\documentclass[aps,prd,twocolumn,superscriptaddress,amsfonts,amssymb,amsmath,eqsecnum,nofootinbib,floatfix]{revtex4}

\usepackage[unicode, colorlinks, citecolor=darkblue, linkcolor=darkred, urlcolor=blue]{hyperref}
\usepackage{graphicx,color,ulem,bm,comment}
\definecolor{darkblue}{rgb}{0,0,0.7}
\definecolor{darkred}{rgb}{0.7,0,0}

\allowdisplaybreaks

\definecolor{dgreen}{rgb}{.3,.7,.3}

\begin{document}
\date{\today}

\title{Constraining temperature distribution inside LIGO test masses from frequencies of their vibrational modes}
\author{Carl Blair}
\email{carl.blair@ligo.org}
\affiliation{Laser Interferometer Gravitational-Wave Observatory (LIGO) Livingston, Louisiana}
\affiliation{Ozgrav, University of Western Australia, Crawey 6009, WA, Australia}

\author{Yuri Levin}
\email{yl3470@columbia.edu}
\affiliation{Center for Theoretical Physics, Department of Physics, Columbia University, New York, NY 10027, USA}
\affiliation{ Center for Computational Astrophysics, Flatiron Institute, New York, NY 10010, USA}
\affiliation{School of Physics and Astronomy, Monash University, Clayton, VIC 3800, Australia}

\author{Eric Thrane}
\email{eric.thrane@monash.edu}
\affiliation{School of Physics and Astronomy, Monash University, Clayton, VIC 3800, Australia}
\affiliation{OzGrav}

\begin{abstract} \noindent
Thermal distortion of test masses, as well as thermal drift of their vibrational mode frequencies, present a major challenge for operation of the Advanced LIGO and Advanced VIRGO interferometers, reducing optical efficiency, which limits sensitivity and potentially causing instabilities which reduce duty-cycle.  In this paper, we demonstrate that test-mass vibrational mode frequency data can be used to overcome some of these difficulties.  First, we derive a general expression for the change in a mode frequency as a function of temperature distribution inside the test mass.  Then we show how the mode frequency dependence on temperature distribution can be used to identify the wavefunction of observed vibrational modes.  We then show how monitoring the frequencies of multiple vibrational modes allows the temperature distribution inside the test mass to be strongly constrained. 
Finally, we demonstrate using simulations, the potential to improve the thermal model of the test mass, providing independent and improved estimates of important parameters such as the coating absorption coefficient and the location of point absorbers.
\end{abstract}

\pacs{interferometers, gravitational waves, elasticity theory, thermal distortion} 
\maketitle
\section{Introduction}
During Advanced LIGO's~\cite{ALIGO} first and second observing runs, about $100\,\hbox{kW}$ of optical laser power was circulated in the Fabry P\'{e}rot arm cavities the interferometers \cite{LIGOsens01}.
During Observation Run 3, 200-250kW was circulating in the arm cavities \cite{O3LIGODetector}. 
It is planned that this power will increase to $0.5 -1$MW \cite{AdvLIGOdes}. 
The heating of mirror surfaces of the test masses associated with this circulating power presents a significant technical challenge, since the thermal deformation of mirror surfaces leads to the loss of optical efficiency.
Optical efficiency is reduced by increased scattered light losses from non-uniform absorption on the mirrored surfaces thermally deforming the surface and by reduced optical coupling between cavities as the beam is altered by thermal lensing.  The reduced optical efficiency ultimately leads to a loss of interferometer sensitivity \cite{LIGOAdaptiveOptics}. 
To address this problem, extra heating is applied to the test masses using specially positioned ring heaters and compensation plates. This is done in such a way that the thermal distortion caused by the ring heaters and compensation plates partially compensates that caused by the laser beam \cite{Zhao06, Brooks16}.

The test masses support a large and complex spectrum of vibrational modes in the frequency range 5-100kHz.  The frequency of these modes depends on the temperature distribution inside the test mass.  Some of these modes are the drivers of parametric instability \cite{Braginsky01,Zhao05}, the control of which was limited by thermal transients~\cite{Blair2017, Hardwick2020}.  Therefore it is useful to monitor the 3-dimensional temperature field inside each test mass for optical efficiency and parametric instability control. 

In this paper we show that components of this temperature field can be measured in real time by monitoring the frequencies of multiple vibrational modes of the test masses.
Some effort has already been spent designing and implementing a system that monitors small changes in mode frequencies \cite{Yu17,BlairCthesis}.
It was shown that hundreds of vibrational modes are visible at the interferometer output at quiescent amplitudes.

These measurements can by extension allow estimates of the thermal distortion of the test mass mirror surfaces and distortion in thermo-optic lens in transmission of the test mass. 
Hartmann wavefront sensor \cite{Brooks2007} are currently used to monitor wavefront distortion in the test masses.  The method proposed here compliments wavefront sensors, providing independent parameter estimates and information from the temperature field dimension along the optic axis.  

The plan of the paper is as follows.  In the next section we develop the mathematical formalism for computing the changes in mode frequencies. 
In section~\ref{sec:test} the formalism is tested against a COMSOL~\cite{COMSOL2013} eigen-frequency analysis.  
In section~\ref{sec:Tdist} we show how to use the frequency changes to make inferences about the temperature distribution inside the test masses and we discuss the limitations of these inferences due to symmetries of the test masses. In section~\ref{sec:real} the estimated temperature distribution from a realistic scenario is examined and in section~\ref{sec:PEst} a Bayesian method for refining test mass thermal model parameters is described.

\section{General formalism}\label{sec:formalism}
\subsection{The preamble: linearity}
The changes in the mode frequencies $\delta\omega_i$ are linear functions of the changes in the temperature inside the test mass, $\delta T(\mathbf{r})$ (here and onwards bold-faced letters denote three-dimensional vectors). Mathematically this can be expressed as follows:
\begin{equation}
\delta\omega_i=\int \rho(\mathbf{r}) f_i(\mathbf{r}) \delta T(\mathbf{r}) d^3\mathbf{r},
\label{omegai}
\end{equation}
where $\rho(\mathbf{r})$ is the density, and functions $f_i(\mathbf{r})$ are form factors that will be discussed in the next subsection. It is convenient to introduce an inner product between functions,
\begin{equation}
\langle f, g\rangle\equiv\int \rho(\mathbf{r}) f(\mathbf{r}) g(\mathbf{r}) d^3\mathbf{r}
\label{innerproduct}
\end{equation}
and similarly between vector fields:
\begin{equation}
\langle \mathbf{a},\mathbf{b}\rangle\equiv\int \rho(\mathbf{r}) \mathbf{a}(\mathbf{r})\cdot\mathbf{b}(\mathbf{r}) d^3\mathbf{r}.
\end{equation}
The factor $\rho(\mathbf{r})$ ensures that the integral is restricted to the test mass volume, and as will be seen below, is useful for expressing orthogonality relations between the test mass mode displacements. Equation (\ref{omegai}) can be written simply as 
\begin{equation}
\delta\omega_i=\langle f_i,\delta T\rangle.
\label{deltaomega}
\end{equation}.

\subsection{Computation of the formfactors \texorpdfstring{$f_i(\mathbf{r})$}{}}
Consider a vector Langrangian displacement $\mathbf{\xi}(\mathbf{r},t)$ of the test mass from its position of rest. In the linear approximation, the elasto-dynamic equations of motion are
\begin{equation}
\rho(\mathbf{r})\frac{\partial ^2\mathbf\xi}{ \partial t^2}=\hat{L}(\mathbf{\xi})
\end{equation}
where $\hat{L}$ is the operator representing the elastic restoring force and given by
\begin{equation}
\hat{L}(\mathbf{\xi})_m\equiv{\frac{\partial\sigma_{mn}}{\partial x_n}}={\frac{\partial \left[c_{mnkl} \epsilon_{kl}\right]}{\partial x_n}},
\end{equation} 
where 
\begin{equation}
\epsilon_{kl}=(\xi_{k,l}+\xi_{l,k})/2 
\end{equation}
is the shear tensor,  $c_{mnkl}$ is the elasticity tensor, and 
\begin{equation}
\sigma_{mn}=c_{mnkl}\epsilon_{kl}
\end{equation}
is the elastic stress tensor. Here the Einstein convention of summing over the repeating tensorial indices is assumed. A normal mode with angular frequency $\omega_i$ is characterized by the  wavefunction $\mathbf{\xi}^{(i)}(\mathbf{r})$ that satisfies the following eigenequation:
\begin{equation}
\hat{L}\left(\mathbf{\xi}^{(i)}\right)=-\omega_i^2\rho(\mathbf{r})\mathbf{\xi}^{(i)}
\label{eigenequation1}
\end{equation}
Importantly, the normal modes satisfy orthogonality relation
\begin{equation}
\langle \mathbf{\xi}^{(i)},\mathbf{\xi}^{(j)}\rangle=\langle \mathbf{\xi}^{(i)},\mathbf{\xi}^{(i)}\rangle \delta_{ij}.
\label{orthogonality1}
\end{equation}

Consider now a perturbation $\delta \hat{L}$ due to the change in temperature of the test mass:
\begin{equation}
\delta \hat{L}(\mathbf{\xi})_m={\partial\over \partial x_n}\left[\delta T(\mathbf{r}){\partial c_{mnkl}\over \partial T}\epsilon_{kl}\right].
\end{equation}
Here we take into account the fact that the elasticity tensor is temperature-dependent. Strictly speaking, there is another contribution to the change in $\hat{L}$ that is due to thermal expansion of the test mass. However, the thermal expansion coefficient LIGO test mass substrate is very small.
Numerically, it is about $0.003\times d \log E/dT$, where $E$ is the Young modulus of the fused silica test mass substrate. Therefore we safely neglect the thermal expansion effect as subdominant.

Consider now the first order perturbation theory for Eq.~(\ref{eigenequation1}), with the perturbed elasticity operator $\hat{L}+\delta\hat{L}$ and perturbed normal-mode wavefunctions $\mathbf{\xi}^{(i)}+\delta\mathbf{\xi}^{(i)}$. This gives
\begin{equation}
\delta \hat{L}\left(\mathbf{\xi}^{(i)}\right)+\hat{L}\left(\mathbf{\delta\xi}^{(i)}\right)=-\omega_i^2\rho(\mathbf{r})\mathbf{\delta\xi}^{(i)}
-2\omega_i\delta\omega_i\rho(\mathbf{r})\mathbf{\xi}^{(i)}.
\label{firstorder}
\end{equation}
By requiring that the perturbed eigenfunction has the same norm as the unperturbed one, we impose an extra constraint
\begin{equation}
\langle \mathbf{\xi}^{(i)}, \mathbf{\delta \xi^{(i)}}\rangle=0.
\label{orthogonality2}
\end{equation}
We now multiply Eq.~(\ref{firstorder}) by $\mathbf{\xi}^{(i)}$, and integrate over the volume.  Using the orthogonality relations Eqs~(\ref{orthogonality1}) and (\ref{orthogonality2}), and the self-adjointness of $[1/\rho(\mathbf{r})]\hat{L}$, we get
\begin{equation}
\langle \mathbf{\xi}^{(i)}, {1\over \rho(\mathbf{r})}\delta \hat{L}\left(\mathbf{\xi}^{(i)}\right)\rangle=-2\omega_i\delta\omega_i\langle \mathbf{\xi}^{(i)}, \mathbf{\xi}^{(i)}\rangle.
\end{equation}
Therefore, the change of the mode's angular frequency is given by
\begin{equation}
\delta \omega_i=-{1\over 2\omega_i}{\langle \mathbf{\xi}^{(i)}, [1/\rho(\mathbf{r})]\delta \hat{L}\left(\mathbf{\xi}^{(i)}\right)\rangle \over \langle \mathbf{\xi}^{(i)}, \mathbf{\xi}^{(i)}\rangle}.
\label{deltaomega1}
\end{equation}

We are now ready to determine the formfactor $f_i(\mathbf{r})$. To acheive this, we write down the numerator of the above equation explicitly as an integral over volume:

\begin{eqnarray}
\langle \mathbf{\xi}^{(i)}, [1/\rho(\mathbf{r})]\delta \hat{L}\left(\mathbf{\xi}^{(i)}\right)\rangle& =&\int \mathbf{\xi}^{(i)}_m \delta\hat{L}\left(\mathbf{\xi}^{(i)}\right)_m d^3\mathbf{r}\nonumber\\
&=&\int \mathbf{\xi}^{(i)}_m {\partial\over \partial x_n}\left[\delta T(\mathbf{r}){\partial c_{mnkl}\over \partial T}\epsilon^{(i)}_{kl}\right] d^3\mathbf{r}\nonumber\\
&=&-\int \delta T(\mathbf{r}){\partial c_{mnkl}\over \partial T}\epsilon^{(i)}_{mn}\epsilon^{(i)}_{kl} d^3\mathbf{r}\nonumber
\end{eqnarray}
The last step is obtained by integrating by parts, using Gauss' theorem, recalling that $\sigma_{mn}=\delta\sigma_{mn}=0$ at the surface of the test mass, and using the symmetry of the elasticity tensor with respect to the indices $m$ and $n$ (the latter insures that the stress tensor is symmetric). From this expression, we conclude that the formfactor is given by
\begin{equation}
f_i(\mathbf{r})={1\over N_i\rho(\mathbf{r})}{\partial c_{mnkl}(\mathbf{r})\over \partial T}\epsilon^{(i)}_{mn}(\mathbf{r})\epsilon^{(i)}_{kl}(\mathbf{r}),
\label{formfactor2}
\end{equation}
where the normalization factor is given by
\begin{equation}
N_i=2\omega_i \int \rho(\mathbf{r}) \left|\mathbf{\xi}^{(i)}(\mathbf{r})\right|^2 d^3\mathbf{r}={4E_i\over \omega_i},
\label{normalization}
\end{equation}
where $E_i$ is the total energy of the mode.
It is worth noting that 
\begin{equation}
c_{mnkl}\epsilon_{mn}\epsilon_{kl}=2U(\mathbf{r})
\label{uelastic}
\end{equation}
where $U(\mathbf{r})$ is the energy density of elastic deformation. For an isotropic medium such as fused silica glass, 

\begin{equation}
c_{mnkl}\epsilon_{mn}\epsilon_{kl}=2U(\mathbf{r})=Y\left(\epsilon_{ll}\right)^2+
2\mu\epsilon^s_{ik}\epsilon^s_{ik},
\label{isotropic}
\end{equation}
where 
$Y$ is the Young modulus, $\mu$ is the shear modulus, and $\epsilon^s_{ik}$ is the incompressible part of the shear,
\begin{equation}
\epsilon^s_{ik}=\epsilon_{ik}-{1\over 3}\epsilon_{ll}\delta_{ik}.
\label{incompressible}
\end{equation}
A simple way of rewriting the formfactor in Eq.~(\ref{formfactor2}), that may be handy in the context of using materials engineering packages like COMSOL or ANSYS, is as follows:
\begin{equation}
f_i(\mathbf{r})={\omega_i\over 2 E_i\rho(\mathbf{r})}\left[{\partial U^{(i)}(\mathbf{r})\over\partial T}\right]_{\xi^{(i)}},
\label{formfactor3}
\end{equation}
where the notation implies that the partial derivative with respect to temperature is evaluated with the mode displacement $\mathbf{\xi}^{(i)}(\mathbf{r})$ being held constant. This completes our computation of the formfactors.



\section{Numerical test and a proposal for practical mode identification}\label{sec:test}
\begin{figure}[ht!]
\centering
\includegraphics[width=0.85\linewidth]{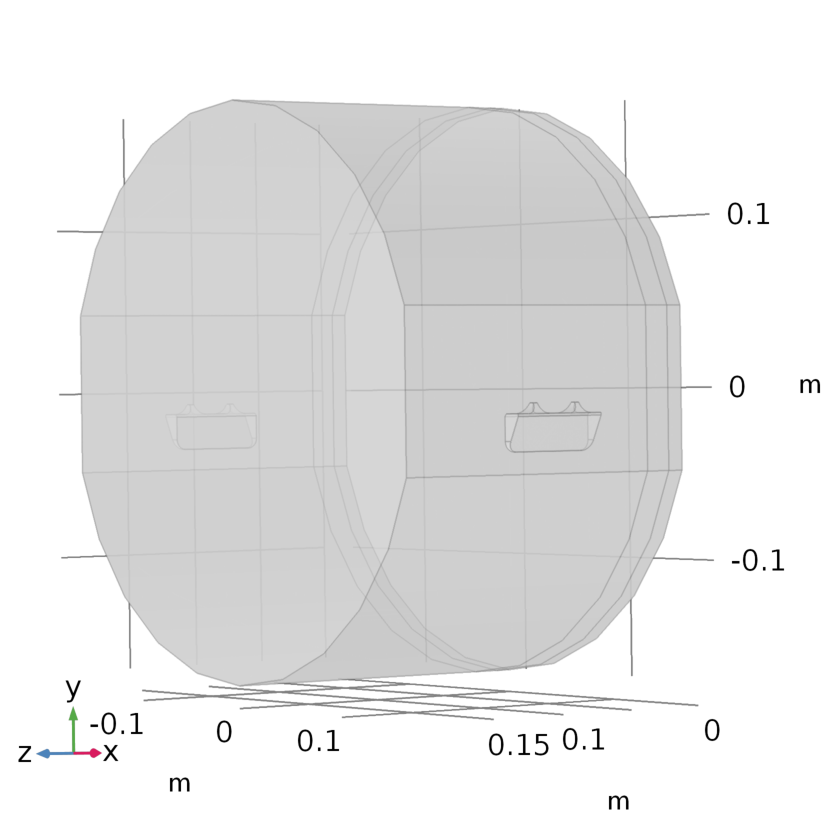} \vspace{-3mm}
\caption[The geometry used for COMSOL simulation]{The geometry used for the COMSOL simulation} \label{fig:COMSOLGeom}
\end{figure}

To validate the form factor solution of Eq.~(\ref{formfactor3}) it is compared to a finite element model eigen-frequency analysis performed with COMSOL \cite{COMSOL2013}. 
The model used is that of the Advanced LIGO test mass.  
Model parameters are given in Table~\ref{tab:ModelParams} and the model geometry is displayed in Figure~\ref{fig:COMSOLGeom}

\begin{table}[ht!]
\caption {Parameters for the COMSOL Model} \label{tab:ModelParams}
\begin{tabular}{ p{2cm}p{2cm}p{4cm}  }
 \hline
 Parameter  & Value & Description \\
 \hline
Diameter    &		340.13mm	&	Diameter \\
Depth       &		199.59mm	&	Depth\\
$\rho$      &		2203kg/$m^3$	&	Density (mass 39564g) \\
Wedge       &		0.07deg	    &	Optic wedge \\
E           &		72.7GPa	    &	Young's modulus \\
$\sigma$    &		0.164	    &	Poisson ratio\\
$\frac{\partial{E}}{\partial{T}}$ &		 11.5\,MPa/K	& Thermal dependence of Young's modulus \\
$\frac{\partial{\sigma}}{\partial{T}}$ & 1.55 $\times 10^{-5}$/K 	& Thermal dependence of Poisson ratio\\
 \hline 
\end{tabular}
\end{table}

The form factors are calculated based on the strain distribution $U^{(i)}$ and total energy $E_i$ of a COMSOL eigen-frequency analysis of the test mass in the ambient (291K) temperature thermal state.  
Form factors and the total mode displacement are shown in Figure~\ref{fig:formfactors_scale} in Appendix A for a selection of modes.

\begin{figure}[ht!]
\centering
\includegraphics[width=0.95\linewidth]{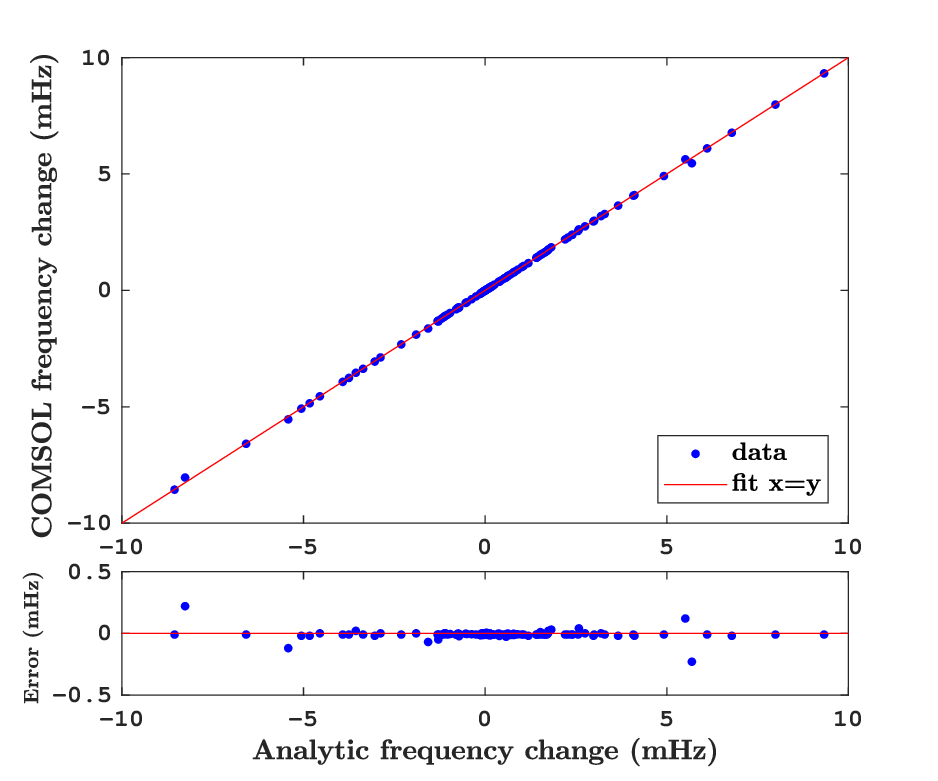} \vspace{-3mm}
\caption{Comparison between the frequency shift predicted by a COMSOL eigen-frequency simulation and the frequency shift predicted by the analytic expression for 225 eigen-modes influenced by an arbitrary thermal disturbance. Excellent agreement is observed.} \label{fig:COMSOLAnalytComp}
\end{figure}

An analytically described change in temperature distribution is defined for the purpose of this test. 
The change in temperature distribution is defined by Zernike polynominal $\mathbb{Z}^3_1$ across the circular surface of the test mass and a uniform distribution through the depth (z) of the test mass.  
Two eigen-frequency analyses are run in COMSOL, one at an ambient temperature of 291K and the second with the additional change in temperature distribution to produce a two sets of eigen-frequencies.
In conjunction, Eqs~(\ref{deltaomega}) and (\ref{formfactor3}) are used to calculate  the expected change in eigen-frequency for that same temperature distribution.  
The results in Figure~\ref{fig:COMSOLAnalytComp} show very good agreement between the analytical expression and the COMSOL simulation.

\begin{figure}[ht!]
\centering
\includegraphics[width=0.95\linewidth]{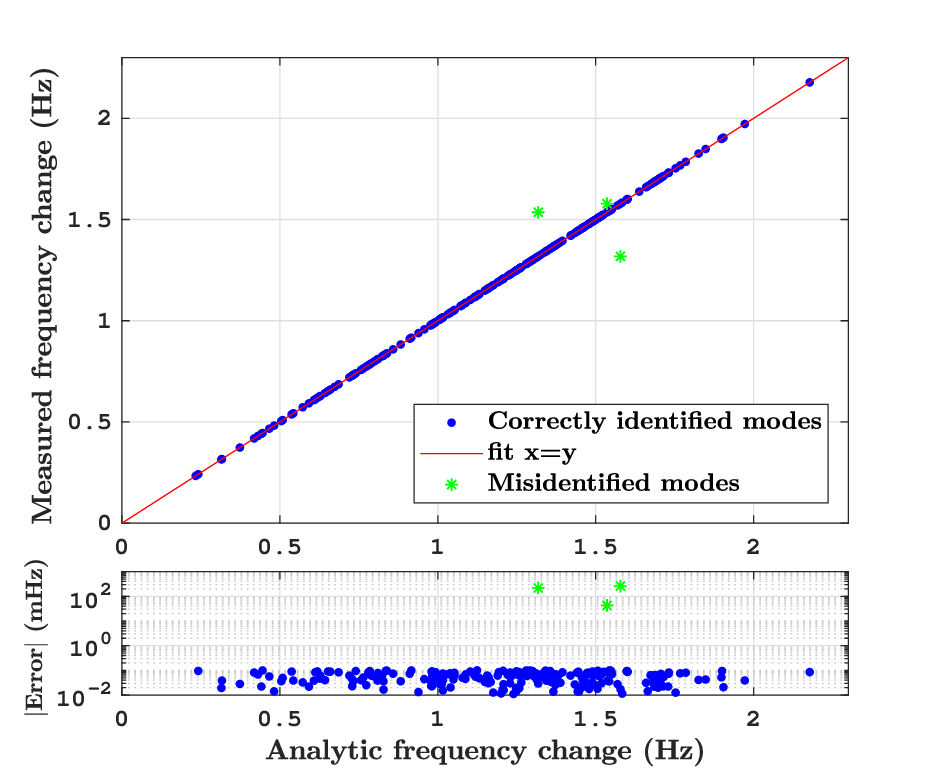} \vspace{-3mm}
\caption{The frequency shift predicted by the analytic expression for 225 eigen-modes influenced by 1\,W of ring heater power plotted against = a simulated measurement including 0.1\,mHz noise and 3 modes misidentified (green) compared to the expectation (red)} \label{fig:ModeIdentification}
\end{figure}

The identification of the mode shape of observed resonances presents a challenge.  
Parametric instabilities \cite{Braginsky01} of vibrational modes with frequencies as high as 47.5\,kHz have been observed at LIGO \cite{BlairCthesis}. 
At this frequency the mode density is high, resulting in several candidate modes that could potentially be causing the instability.  
Furthermore, theoretical calculations show that with increased circulating power there might be instabilities caused by modes with frequencies as high as 90\,kHz~\cite{Gras10}.
(The recent installation of acoustic mode dampers \cite{Biscans19} makes high frequency instabilities a lot less likely.) 
Knowledge of the mode shape is required to design active control schemes that apply forces to the test masses \cite{Blair2017} or optical feedback \cite{Fan2010} \cite{Bossolkov2020}. 
Currently modes are identified by comparing observed resonant frequencies with those computed using finite element modelling.
Confident mode identification is currently limited to 17\,kHz.  
At higher frequencies, imprecise knowledge of the elasticity parameters of fused silica produce large enough errors such that confusion between modes is a serious issue.  
The analytical expression for change of mode frequencies as a function of temperature presented here presents a new tool for mode identification.  
This could work as follows: 
1. a well-controlled thermal transient perturbation is applied to the test mass, and the internal temperature distribution is computed as a function of time using finite element modelling. 
2. The transient change in mode frequencies can be calculated as a function of time using the formalism presented above or finite element modelling. 
3. These are compared and matched to the measured transient frequency changes in the monitored modes.  
By following this procedure,  mode identifications can be confirmed.  
A simulated example of such a confirmation is shown in Figure~\ref{fig:ModeIdentification} where 3 modes (colored green) have been deliberately misidentified by switching their indices.  
Simulated mode frequencies on the vertical axis are compared with mode frequencies calculated with the analytical expression of Eq.~(\ref{omegai}).  
In this case the temperature field is simulated in COMSOL as the steady state for 1W applied ring heater power.  
The COMSOL simulation includes thermal expansion and a uniform 0.1\,mHz measurement noise has been added to the COMSOL simulated eigen-frequencies. 
The three misidentified modes can be clearly identified as outliers.
The correct mode identification is critical for active control of parametric instability and is also required to make inferences about the temperature distribution from measurements of the eigen-frequencies of the test mass. 

\section{Constraining the temperature field inside the test mass}\label{sec:Tdist}

One might suppose that if one is able to measure the temperature-induced frequency shifts of all of the vibrational modes to arbitrary precision, one should be able to reconstruct the 3-dimensional temperature perturbation inside the test mass. 
This would be an unprecedented fit for experiments with solids as far as we know. 
However, as we explain below, this strategy runs into problems because the form-factors $f_i(\bf{r})$ do not necessarily form a complete basis for all of the possible temperature perturbations; we show this explicitly for the case when the test mass has a reflection symmetry, as they in fact, do. 
We begin however in the next subsection by considering the conceptually simple case where the formfactors do form a complete basis and the temperature perturbation can, in principle, be measured.

\subsection{Case of \texorpdfstring{$f_i({\bf r})$}{form factors} forming a complete basis}.

Completeness allows us expand $\delta T(\mathbf{r})$ in a series:
\begin{equation}
\delta T(\mathbf{r})= p_i f_i(\mathbf{r}).
\end{equation}
Here we use the Einstein convention, where the summation of repeated indices is assumed.
In general, one expects the functions $f_i(\mathbf{r})$ to be linearly independent, however in some cases where a high degree of symmetry is present, it may turn out that this is not so. In such situation, one needs to restrict  the series above to a linearly independent subset of functions spanning the whole function space, so that the expansion is unique.

Substituting the expansion above into Eq.~(\ref{deltaomega}) results in a matrix equation
\begin{equation}
\delta \omega_i=C_{ij}p_j,
\end{equation}
where 
\begin{equation}
C_{ij}\equiv \langle f_i,f_j\rangle.
\label{cmatrix}
\end{equation}
One therefore has 
\begin{equation}
\delta T(\mathbf{r})=C^{-1}_{ij} \delta\omega_i f_j(\mathbf{r}),
\label{expansion}
\end{equation}
where $C^{-1}_{ij}$ are the elements of $C^{-1}$. Since one monitors only finite amount $N$ of the normal modes, in practice one should restrict $C_{ij}$ to be the $N$-dimensional square matrix.

\subsection{Incompleteness of \texorpdfstring{$f_i({\bf r})$}{form factors} due to symmetry of the test mass}

We do not in fact have a mathematical proof that the set of functions $f_i({\bf r})$ is ever 
complete for a generic shape of the test mass, although intuitively it seems likely. However,
a practically important counterexample is the case when the test mass has a reflection symmetry, say $z\rightarrow -z$. In this case the vibrational modes have either odd or even parity with respect to $z$, but because it is the elastic energy density that determines the calculations
of the formfactors in Eq.~(\ref{formfactor3}), 

\begin{equation}
f_i(x,y,z)=f_i(x,y,-z),
\label{simmetry1}
\end{equation}

i.e. the form factors all have even parity, see Fig.~\ref{fig:formfactors_scale}. Therefore the frequency shifts will carry no information about the odd part of the temperature perturbation, 

\begin{equation}
    \delta T_{\rm odd}(x,y,z)={1\over 2}\left[\delta T(x,y,z)-\delta T(x,y,-z)\right],
\end{equation}
but will instead only carry information about the even part of the temperature perturbation
\begin{equation}
    \delta T_{\rm even}(x,y,z)={1\over 2}\left[\delta T(x,y,z)+\delta T(x,y,-z)\right].
\end{equation}

\begin{figure}[ht!]
\centering
\includegraphics[width=0.95\linewidth]{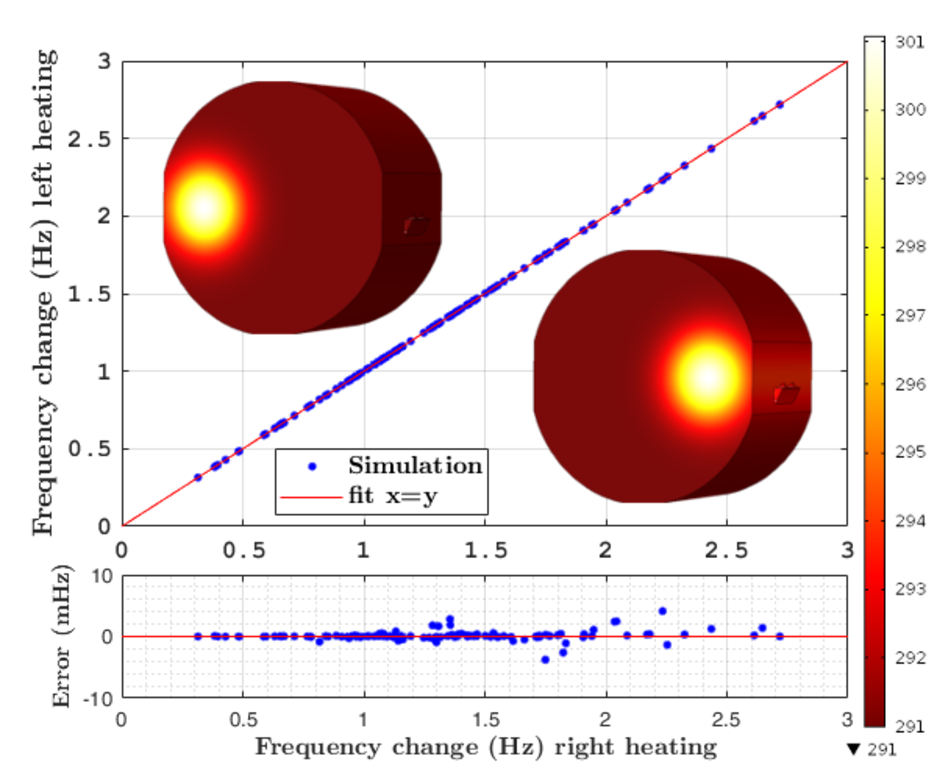} \vspace{-3mm}
\caption{Comparison of the mode frequency shift between two thermal profiles (inset) that are symmetric left to right.  The mode frequencies of the case where the heating is to the left of the optic axis (vertical axis) are indistinguishable from the mode frequencies where the heating is to the right of the optic axis (horizontal axis)} \label{fig:ThermSym}
\end{figure}

There are three approximate reflection symmetries in the LIGO test masses that result in degeneracy.  Symmetry front to back $z\rightarrow -z$, the symmetry right to left $x\rightarrow -x$ and symmetry up and down $y\rightarrow -y$.  The degeneracy associated with these symmetries result in two thermal profiles that are related by one of these symmetries, being indistinguishable. As a practical illustration,
in Figure~\ref{fig:ThermSym} we show two different thermal profiles as well as  the  change in mode frequencies computed in COMSOL for each of the profiles.

The thermal profiles have intentionally been selected to have symmetry right to left.  They are both 2D gaussian profiles across the mirror surface, uniform in depth.  
As expected the changes are almost equal, with precision of approximately 1\%.  

With the approximate symmetries of the Advanced LIGO test mass the maximum information that can be inferred from the frequency shifts is the symmetrized temperature distribution defined on one octant of the test mass:
\begin{equation}
    \delta T_{\rm sym}(x,y,z)=(1/8){\Large \Sigma}\delta T[\pm x, \pm y, \pm z]
\end{equation}
for $x,y,z>0$, here $\Sigma$ denotes the summation over all possible combination of signs of $x,y,z$ and the origin is assumed to be located at the center of mass of the test mass.

There may be a way of breaking some of the degeneracy by measuring other temperature-sensitive observables such as the distortion of the mirror's surface, or the thermal lensing of light passing through the test mass, for example with the Hartmann sensor. However, we do not consider these possibilities any further and leave their consideration for future work.

\subsection{\texorpdfstring{$3$-}{Three }dimensional temperature reconstruction using Singular Value Decomposition }

Suppose now that we are considering properly symmetrized temperature fields so that $f_i({\bf r})$ do form a complete basis. We should still exercise caution using Eq.~(\ref{expansion}) for the temperature field reconstruction. 
Similarity between some formfactors $f_i({\bf r})$ means the $C$ matrix is ill-conditioned (one or more eigenvalues are close to zero), the inversion becomes numerically unstable, leading to unreliable results. 
A common way of dealing with ill-conditioned matrices is to regularize the matrix by singular-value decomposition (SVD).   
The conversion matrix is decomposed into orthogonal matrix $\mathbf{U}$, diagonal matrix $\mathbf{C'}$ and another orthogonal matrix $\mathbf{V}$.
\begin{equation}
\mathbf{C}=\mathbf{U}\mathbf{C}'\mathbf{V}^*,
\label{svd}
\end{equation}
We adopt the convention that $\mathbf{C}'$ is defined with values sorted from largest to smallest along the diagonal.
Since $C$ is real, the Hermitian transpose can be replaced by a regular transpose $\mathbf{V}^*=\mathbf{V}^T$.
By removing eigenmodes associated with small eigenvalues, we reduce the dimensionality of $C'$ to $N-\alpha$ by removing the $\alpha$ smallest elements of the complete diagonal matrix $\mathbf{C}'$ and the $\alpha$ associate eigenvectors in $\mathbf{U}$ and $\mathbf{V}$.
If $\alpha$ is suitably chosen, the resulting ``regularized'' matrix is numerically invertible.
In what follows an example of singular value decomposition applied to simulated eigen-frequency data is demonstrated.  

\begin{figure}[htbp]
\centering
\includegraphics[width=0.95\linewidth]{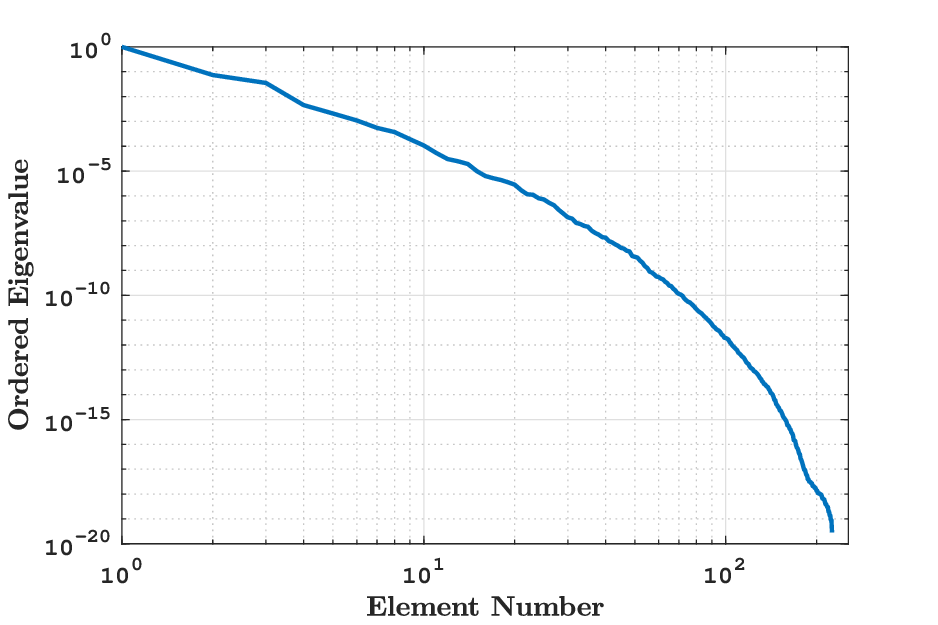} \caption[Eigenvalues of the SVD matrix C`.]{Eigenvalues of the SVD matrix C'} \label{fig:svdEigenValues}
\end{figure}

The form factors for the first 225 eigenmodes of the test mass are calculated in COMSOL, each form factor is defined by the 60000 vertex elements existing in the three dimensional domain of the test mass.  
The inner product defined in Eq.~(\ref{innerproduct}) is performed to determine the conversion matrix $\mathbf{C}$.  
Then the singular value decomposition is performed with Eq.~(\ref{svd}).  
The relative numerical value of the eigenvalues (diagonal elements of $\mathbf{C}'$) provides a measure of the additional information that can be recovered by adding each new element in the SVD.  
These values are plotted in Figure~\ref{fig:svdEigenValues}.  
From the figure it can be seen that using more than 100 SVD elements does not provide a significant increase in information.  
It is also interesting to consider the shape of the largest elements in $\mathbf{C}'$ as these represent the temperature distribution components that will be most easily recovered.  
A selection of $\mathbf{C}'$ eigenfunctions are shown in Appendix~\ref{sec:SVDEle}. 

\begin{figure}[tbh]
\centering
\includegraphics[width=0.95\linewidth]{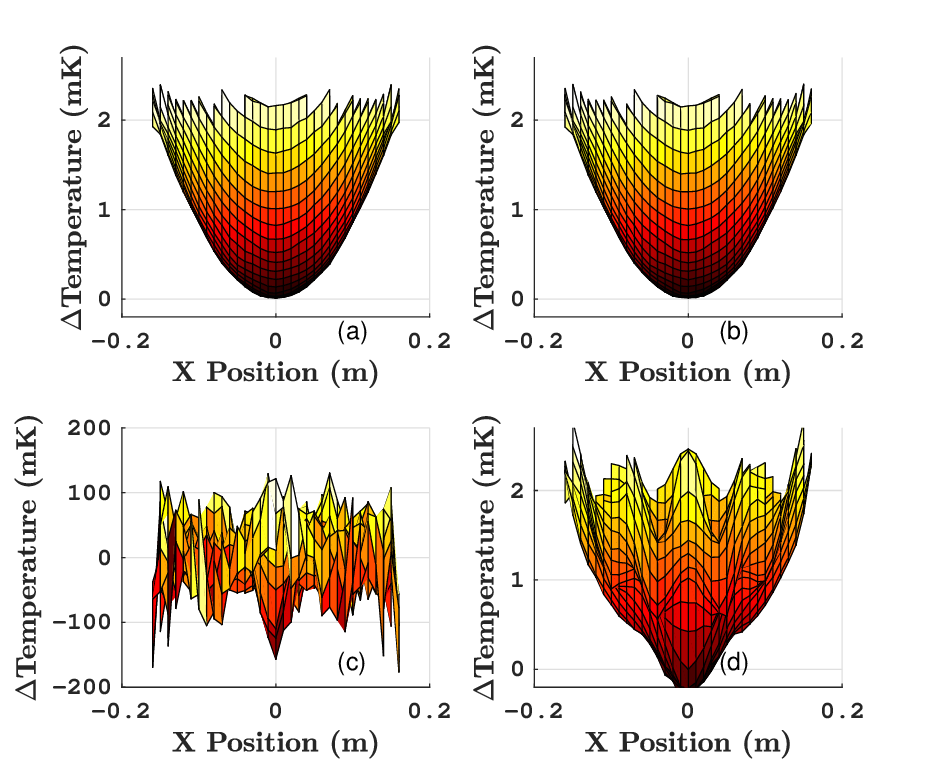} 
\caption[SVD inversion]{Profile of estimated temperature distribution inferred from changes in eigenfrequecy a) Inverting $C_{ij}$ directly, b) SVD inversion using all 225 eigenfrequencies, c) SVD inversion using all 225 with 0.1mHz noise, d) SVD inversion using 86 components with 0.1mHz noise.} \label{fig:svdExample}
\end{figure}

To demonstrate the usefulness of SVD, we consider a rotationally symmetric temperature distribution
\begin{equation}
 T=T_0+dT=T_0+(1/12)r^2 , 
 \label{eq:tdist1}
\end{equation}
where r is the radial cylindrical coordinate and $T_0$ is a constant, 
and compute using Eqs.~(\ref{omegai}) and (\ref{formfactor3}) the changes in the $225$ test mass eigenfreqencies.  
The temperature distribution possesses all the required symmetries and can thus be recovered by inverting the conversion matrix, with or without using the SVD.  
The corresponding temperature profiles are shown in Figures \ref{fig:svdExample} (a) and (b), without any significant difference in quality.  
However if we now assume that the eigenfrequency measurements are not perfect and contain errors, we note a marked difference in the quality of reconstructed temperature fields. As an example, we add a random frequency error drawn from a normal distribution with width $0.1$\,mHz to each analytically computed eigen-frequency change, and then compute the temperature fields from this erroneous data set.  

We observe that the truncated SVD inversion in Figure \ref{fig:svdExample} (d) produces a significantly better result compared to the inversion that uses matrix $C$ directly in Figure \ref{fig:svdExample} (c). The latter is distorted due to errors in poorly resolved eignmodes.

\begin{figure}[t!]
\centering
\includegraphics[width=0.95\linewidth]{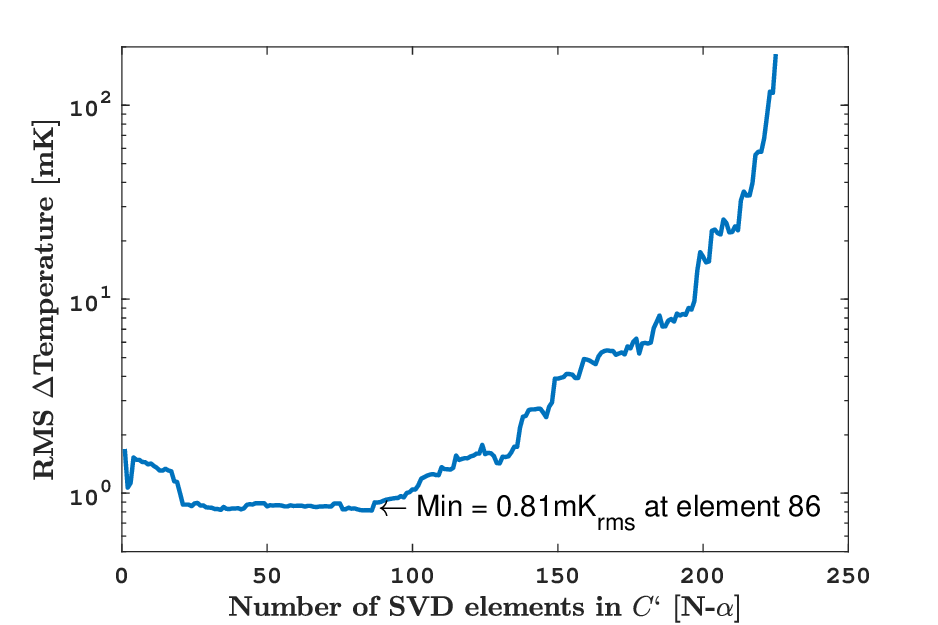} 
\caption[RMS error of estimated temperature]{RMS error of estimated temperature as a function of the number of singular value decomposition elements used in the temperature reconstruction.  The minimum error occurs with 86 elements for this particular temperature distribution.} \label{fig:rmsError}
\end{figure}

The optimal choice for $\alpha$, the number of excluded eigenvectors, can be estimated for any particular temperature distribution, with a particular noise distribution by comparing the rms error of the temperature estimate
\begin{align}
b_{rms}=\oint_V{|{dT-\delta T}|}dr .
\end{align}
In figure \ref{fig:rmsError}, the example plot of rms error as a function of the number of SVD elements used is shown.  
This example uses the same data as Figure \ref{fig:svdExample}.
The optimal number of SVD elements is 86.  The reconstructed temperature field with 86 elements is shown in Figure  \ref{fig:svdExample} d. 

\section{Realistic Temperature Distributions}\label{sec:real}
\begin{figure}[htbp]
\centering
\includegraphics[width=0.95\linewidth]{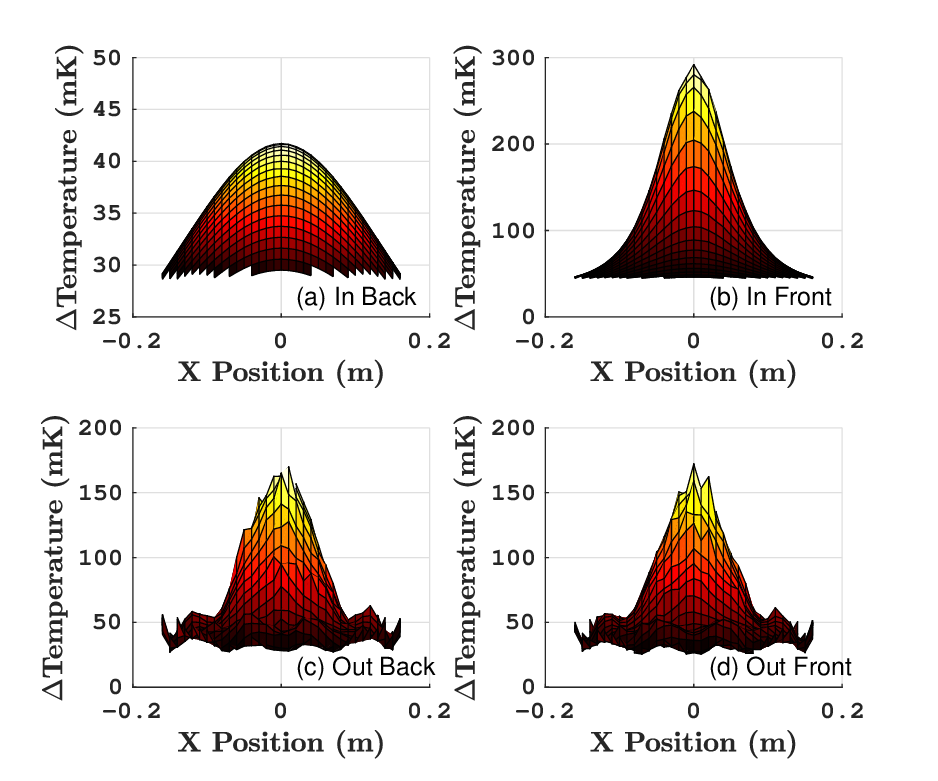}
\caption[Input and reconstructed temperature distribution]{Temperature distribution of (a) back surface and (b) front surface and reconstructed temperature distributions from (c) back surface and (d) front surface of the optic. Reconstruction is done using singular value decomposition  using 81 elements and negligible (1nHz) measurement noise.} \label{fig:asymReconstruct}
\end{figure}

\begin{figure}[h!]
\centering
\includegraphics[width=0.95\linewidth]{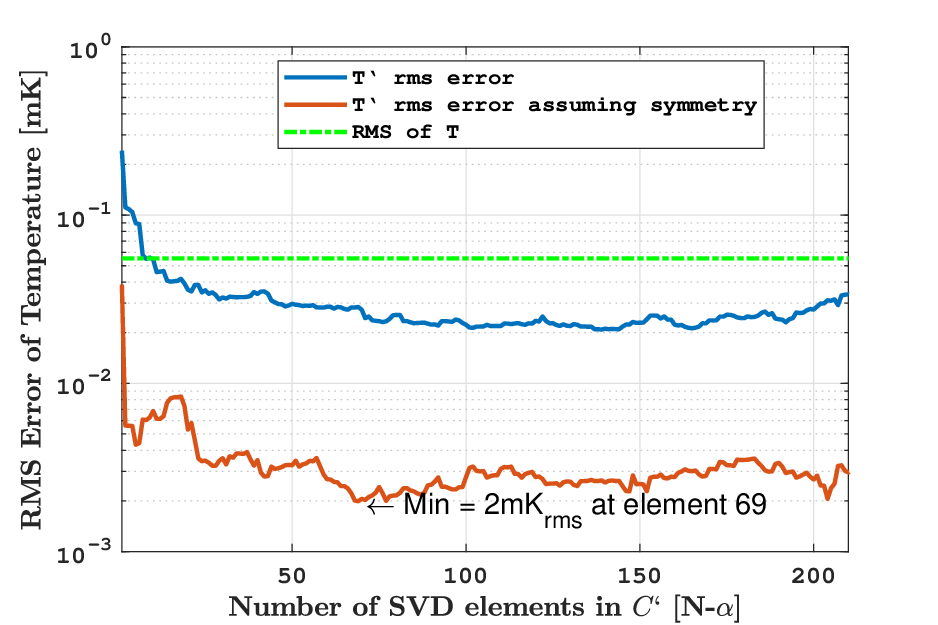}
\caption[RMS error of asymmetric temperature distributions]{Comparison of RMS error of asymmetric temperature distributions assuming a symmetric (tan) and standard (blue) model.  This is compared to the RMS of the temperature distribution decomposed into singular value decomposition elements (red), and the RMS of the temperature distribution (green dot)} \label{fig:rmsSymmetry}
\end{figure}

It is useful to reconstruct a temperature field that does not possesses all the symmetries of the test mass in order to see how our analysis works in real-world conditions.  
In this section we demonstrate that the recovery of symmetrized temperature distribution is indeed possible, circumventing the completeness problem due to test mass symmetries.
If the temperature distribution is not symmetric in the same manner as the test mass symmetry the resulting rms error of the temperature distribution is large.  
In Fig.~\ref{fig:rmsSymmetry} this is demonstrated.  
In this case a a 100\,kW beam on an optic with a uniform 1\,ppm coating absorption is simulated resulting in a temperature distribution that is relatively higher on the high reflectivity surface and relatively cooler on the opposing surface (Figure~\ref{fig:asymReconstruct} top left and right panels respectively).  

The estimated temperature distribution from the change in eigen-frequencies has roughly the same temperature distribution on the high reflectivity surface and the opposing surface (Figure~\ref{fig:asymReconstruct} bottom left and right panels respectively).  
However, the average of the front and back surfaces of the estimated temperature distribution is approximately equal to the average of the front and back surfaces of the input temperature distribution.   

This can be appreciated by comparing the rms error of the total test mass temperature distribution (blue) and the rms error of a model that uses half the test mass averaged with a reflection symmetry in the z-axis in (red) Figure ~\ref{fig:rmsSymmetry}, i.e.,
\begin{align}
\delta T_{\rm sym}(x,y,z)=(1/2){\Large \Sigma}\delta T[ x, y, \pm z],
\end{align}
where $z>0$.  
More generally the average temperature distribution over one octant of the test mass may be computed when considering an arbitrary temperature distribution.  
While some information is lost, the symetrized temperature distribution still provides useful information. One potentially useful example is the measurement of the radial position of a beam on a test mass.

\vspace{-5mm}
\section{Parameter Estimation using the 3D Temperature Field}\label{sec:PEst}
\vspace{-3mm}
In the previous sections it was demonstrated measurements of a set of eigen-frequencies can be used to measure temperature distribution.  
The temperature distribution in the optic at LIGO may be defined by a relatively small number of parameters~\cite{Yu17} of a thermal model.  This limited model is described in Table~\ref{tab:ThermalParams}.

\begin{table}[h!]
\caption {LIGO Test-mass Thermal Model Parameters} \label{tab:ThermalParams}
\begin{tabular}{ p{2cm}p{2cm}p{4cm}  }
 \hline
 Parameter  & Value & Description \\
 \hline
 w    &		51.0$\pm$0.1\,mm	&	Beam radius \\
 Y   &		11$\pm$1\,mm	    &	Beam height ref center\\
 X   &		8$\pm$1\,mm	        &	Beam pos ref center\\
 $P_{abs}$ &	0.2$\pm$0.1\,W	&	Power absorbed in coating \\
 k   &		1.38$\pm$0.01 W/m.K	    &	Thermal Conductivity \\
 $\alpha$    &		(0.52$\pm$0.01) $E^{-6}$/K	    &	 Thermal Expansion  \\
 $C_V$         &		703$\pm$10 J/Kg.K	    &	 Specific heat  \\
$\rho$  &		2203$\pm$1 Kg/$m^3$	    &	 Density  \\
 $\epsilon_{fs}$          &		0.9$\pm$0.05 J/Kg.K	  & Emissivity SiO$_2$  \\
 $\epsilon_{coat}$ &		0.9$\pm$0.1 J/Kg.K	  
 & Emissivity coatings  	 \\
 \hline 
\end{tabular}
\end{table}

In this section we show that the measurements of eigenfrequecnies can be used to measure specific thermal model parameters. We demonstrate how this can be done with Advanced LIGO data using a Bayesian approach.  
Eigenfrequency information can be collected during normal Advanced LIGO operation.  
We note that thermal conductivity affects the time evolution of the temperature inside the test mass. 
Therefore rather than using eigenfrequency measurements at one point in time we should use measurements over a time span $\delta \omega_i(t_j)$.  

Some parameters such as laser power absorbed in the mirror coating (from the previous section) affect the temperature distribution in a linear fashion:
\begin{align}
T({P_{abs}})\propto P_{abs} \times T(P_{abs}=1) .
\end{align}
In this case, the power absorbed in the coating $P_\text{abs}$ may be inferred by linear regression:
\begin{equation}\label{eq:rhoEstimate}
     \hat{P}_{abs} = \sum\limits_i{\frac{\delta\omega_i(t_j)\delta\omega_i(t_j|P_{abs})}{2\sigma_i^2}} \Bigg/ \sum\limits_i{\frac{(\delta\omega_i(t_j|P_{abs}))^2}{{2\sigma_i^2}}}
\end{equation}
Previously, a single eigenmode has been used in such an analysis \cite{Yu17}.  
Linear regression using many eigenmodes benefits from more data and therefore less susceptibility to noise.  
Using more than one eigenmode also provides additional robustness against errors in different thermal model parameters.  
As errors produce temperature distributions with components orthogonal to the temperature distribution of interest, only the temperature distribution component common to both model parameters will affect the result. 
For a concrete example, consider a numerical experiment with $P_{abs}$= 0.2\,ppm.
Consider that there is a 10\% error in thermal conductivity such that $k=1.38$ for the calculation of $\delta\omega_i(t_j)$ and $k=1.52$ for the calculation of $\delta\omega_i(t_j|P_{abs})$.
We then compare $\hat{P_{abs}}$ calculated with Equation~\ref{eq:rhoEstimate}
with one eigenfrequency (the 6$^{th}$ mode at 9330\,Hz) and 100 eigenfreuencies (ranging from 5740 to 24888Hz).  
With one eigenfrequency, the estimate is biased and inaccurate $\hat{P_{abs}}$ = - 0.162$\pm$1.5\,ppm.
Using 100 eigenfrequencies, the estimate is precise and accurate $\hat{P_{abs}}$ =  0.199$\pm$0.005\,ppm. 
This demonstrates the significant improvement in accuracy and robustness achieved by using a large number of eigenfrequencies. 

More generally a set of thermal model parameters $\bm{ \Gamma}$ may be estimated by locating the peak in the likelihood function.
\begin{equation}\label{eq:combined_L}
    \log{\cal L}({\bm{{\delta\omega}}} | \bm{\Gamma}) = \sum_i^N \sum_j^M
    -\frac{1}{2 \sigma_i^2}
    \Big(\delta\omega_m^i(t_j) - \delta \omega^i(t_j|\bm{\Gamma})\Big)^2
\end{equation}
In this paper we explore the likelihood function over various parameter spaces to determine what information is most easily recovered using this technique.

Transients in temperature are caused by laser light being absorbed in the test-mass mirror coating, changes in ring heater power and changes in ambient temperature.  In this section, we focus on transients caused by laser light absorbed in the mirror coating as this is the most common thermal transient in LIGO optics.  The thermal model for such a transient in its simplest form is defined by the optic geometry, the material properties of fused silica, and the properties of heat sources defined in Tables~\ref{tab:ModelParams} and \ref{tab:ThermalParams}.  Laser light is absorbed in the mirror surface.  Thermal equilibrium is attained when the radiative cooling to the thermal bath balances the heat load on the mirror surface.

\begin{figure}[hbt!]
\centering
\includegraphics[width=0.95\linewidth]{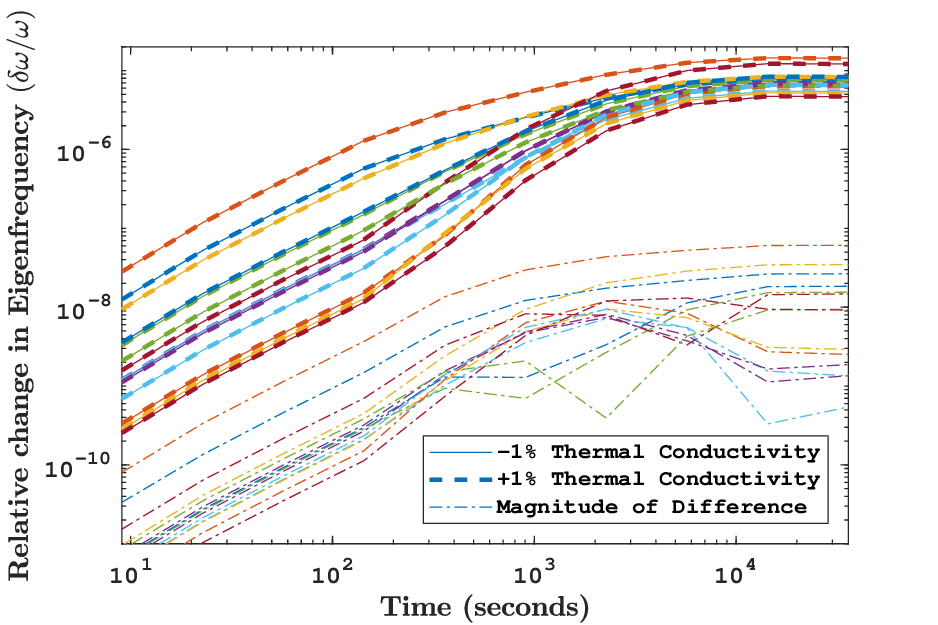} 
\caption[Time evolution of Eigenfrequencies for 2 thermal conductivities]{Time evolution of a selection of eigenfrequencies for thermal conductivity of $k$=1.37 (dashed) and $k$=1.39 (solid) and the difference (dash dot)} \label{fig:ThermalConductivitytime}
\end{figure}

\begin{figure}[hbt]
\centering
\includegraphics[width=0.95\linewidth]{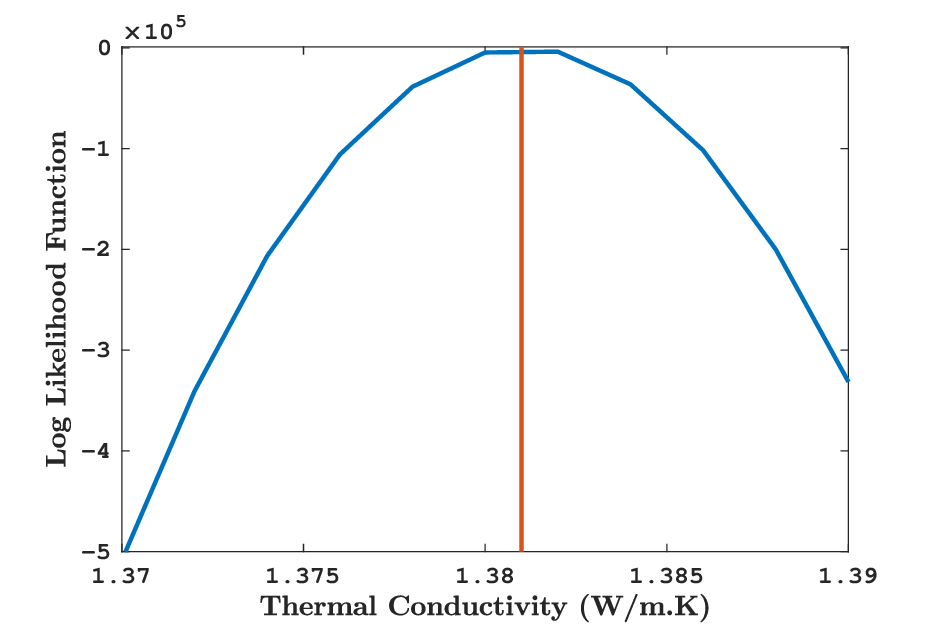} 
\caption[Log likelihood function, thermal conductivity example]{Log likelihood function example to thermal conductivity.  Using the data like that in Figure~\ref{fig:ThermalConductivitytime}, including 100modes and minimal (1\,nHz) measurement noise on the simulated measurement of a test mass with $k$=1.381W/(m.K) thermal conductivity. } \label{fig:ThermalConductivityLL}
\end{figure}

Information recovered from multiple eigen-frequencies represent temperature gradients in the optic.  
Thermal gradients dissipate on a time scale proportional to the thermal gradient length scale.  
Therefore, the timescale of interest depends on characteristic length scale of the expected temperature field.  
For illustrative examples, and to keep computational costs low, we assume the eigenfrequneices are measured 10 times, with $t_i$ logarithmically spaced between 3 seconds and 10 hours.

A simulated example transient of a selection of eigenfrequencies is shown in Figure \ref{fig:ThermalConductivitytime} for two different values of thermal conductivity $k$; the difference between the eigenfrequencies evaluated with different thermal conductivity is shown as a dot-dash line.  
Note that most of the action, where mode frequencies change relative to each other, happens between a few hundred and a few thousand seconds.  
This is therefore the region we would expect to get most information regarding differences in temperature distribution for different thermal conductivity.  

The log likelihood function is calculated for a simulated measurement point of ${k}={1.381}$ and is plotted in Figure \ref{fig:ThermalConductivityLL}.  The log likelihood function peaks at the simulated measurement point, showing that this method can be used to infer properties like the thermal conductivity. 
This example assumes no uncertainty on any other parameters.

\begin{figure}[hbt]
\centering
\includegraphics[width=0.95\linewidth]{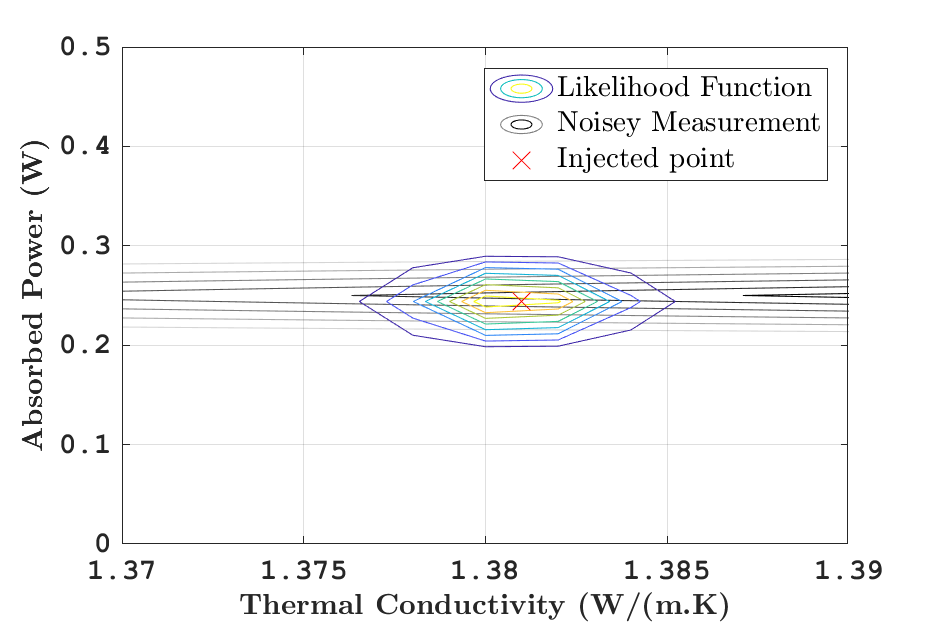} 
\caption[Likelihood function of point absorber power and thermal conductivity]{likelihood function of point absorber power and thermal conductivity with minimal measurement noise (1\,nHz) (coloured contour lines) and with 0.1\,mHz measurement noise (gray contour lines). It can be seen that with measurement noise the absorbed power may be constrained while the thermal conductivity may only be minimally constrained} \label{fig:AbsTCLL}
\end{figure}

As apparent from Table~\ref{tab:ThermalParams} there are many model parameters that are subject to significant uncertainties.  To get a sense of what parameters may be constrained using the method defined in this paper we investigated parameters in pairs.   In Figure~\ref{fig:AbsTCLL}, the likelihood function is plotted over the parameter space of absorbed power $P_{Abs}$ and thermal conductivity $k$.  The injected point is marked red.  It can be seen that with small additional noise (1\,nHz) other than quantization noise of the finite element simulation, both parameters are well constrained (coloured contour lines).  However with 0.1\,mHz measurement noise, the absorbed power is well constrained while the thermal conductivity can not be well constrained (grey lines with values indicated).  

These simulation were done for many pairs of parameters in Table~\ref{tab:ThermalParams}.  Generally the absorbed power and the beam radial position are reasonably well constrained while the $X$ and $Y$ position estimates are less well constrained. Emissivity can be constrained in a similar manner to thermal conductivity.  Other parameters are less well constrained.

\begin{figure}[htb]
\centering
\includegraphics[width=0.95\linewidth]{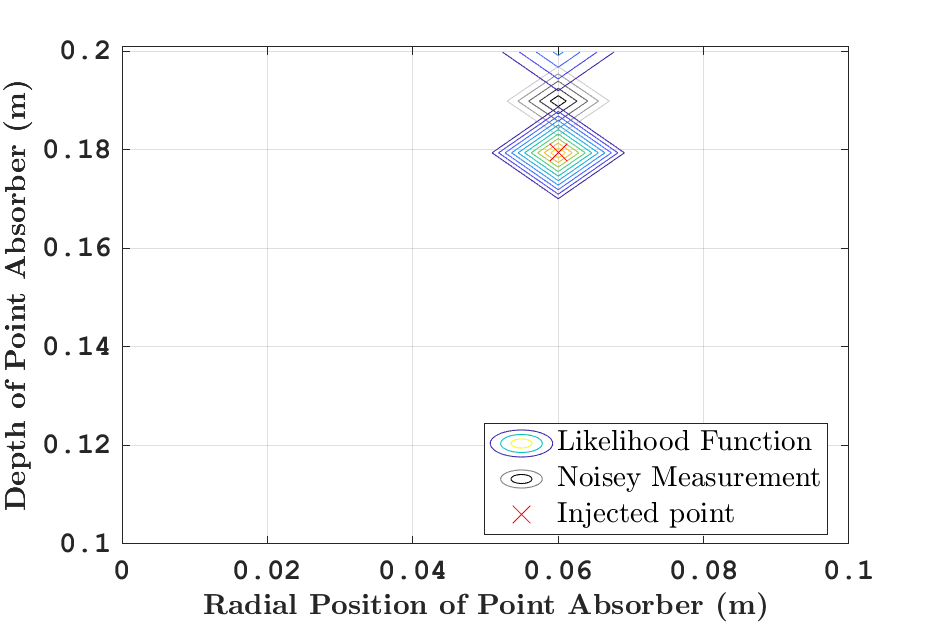} 
\caption[likelihood function of Point absorber power and Thermal conductivity]{likelihood function of point absorber location in radial position and depth  with 5\,nHz measurement noise (in colour). With 0.1\,mHz measurement noise (gray contour lines) a position bias is introduced} \label{fig:PointAbsLocLL}
\end{figure}

Finally, in this section we show how this technique can be used to estimate thermal model parameters that are not accessible with Hartmann wavefront sensors.  
The thermal model of Table~\ref{tab:ThermalParams} assumes uniform absorption in the mirror high reflectivity coating.  
This model has recently been shown to be inadequate \cite{O3LIGODetector}. 
Point absorbers on the high reflectively surface of the test mass produce significant heating.  
The position of such a point absorber can be recovered well with the methods presented here.
However this information is also accessible with the Hartmann wavefront sensor.  
In the following simulation we imagine a situation where instead of coating point absorbers there is a point absorber in the bulk of the test mass. 
The point absorber is a 30\,um, 10\% absorption feature.  Figure~\ref{fig:PointAbsLocLL} shows the likelihood function for the data given the point absorber location along with the true value of the point absorber location in red.  
While the distribution is bimodal, in this particular case, the absorber position is recovered as the maximum in the likelihood function, however with realistic measurement noise a bias is introduced.    

The thermal transient due to change in laser power is a common occurrence happening about once per day.  Therefore a multi-parameter estimation may be arbitrarily refined using a Bayesian approach where the posterior distribution of the thermal model parameters inferred from one transient in laser power is used as the prior distribution for the subsequent measurement.
\vspace{-4mm}
\section{Conclusion }\label{sec:Conc}
\vspace{-4mm}
Establishing robust thermal control of the test masses is one of the important tasks that will allow LIGO and Virgo to attain design goals.  In this paper we provided an efficient method of computation of the vibrational mode frequency response to a temperature perturbation in the test mass.
We demonstrated that the method may be inverted, enabling the conversion of vibrational mode frequency measurements into temperature distribution information.
Finally, it was demonstrated that parameters of the test-mass thermal model may be estimated with improved  precision using this temperature distribution information.  
Symmetries of the test mass prevent the recovery of complete 3D temperature distribution information, only symmetric components of the temperature distribution may be recovered.
In principle, information from the Hartmann sensor could be used to break degeneracy between these symmetries and provide more information on the 3D temperature distribution. 
The framework described in this paper is demonstrated to provide useful coating absorption estimates and may allow estimates of several other thermal model parameters.  However, this is dependent the nature of the measurement noise.
Further experimental work and on-site measurements are needed to determine how the techniques proposed in this paper will be helpful for thermal control of the test masses. 

The initial stages of this research were supported by Levin's Australian Research Council (ARC) Future Fellowship and Blair's PhD scholarship.  Later work was supported by Dr. Blair's Caltech postoctoral fellowship and ARC DECRA DE190100437.
ET is supported through ARC Future Fellowship FT150100281 and ARC Centre of Excellence CE170100004.
\vspace{-4mm}
\bibliography{myRefs}

\appendix
\vspace{-4mm}
\section{Form Factor Distributions}\label{sec:FFDist}
\begin{figure}[hb]
\centering
\includegraphics[width=0.38\linewidth]{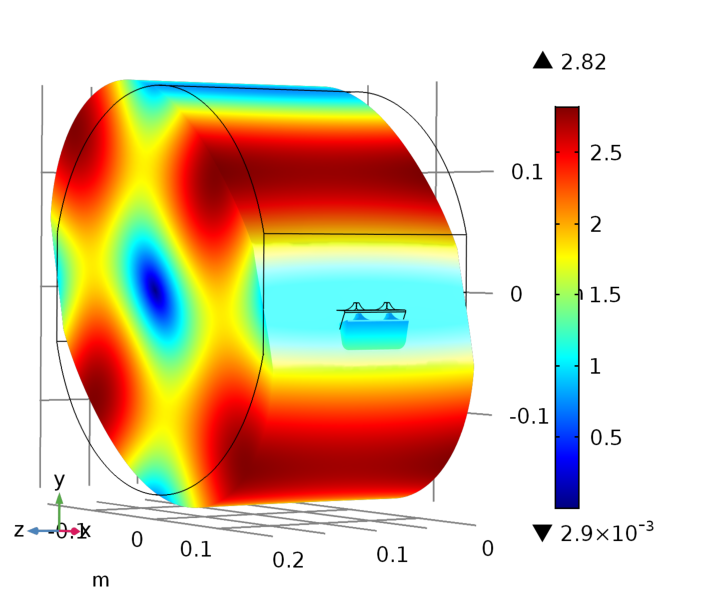} \vspace{-1mm}
\includegraphics[width=0.38\linewidth]{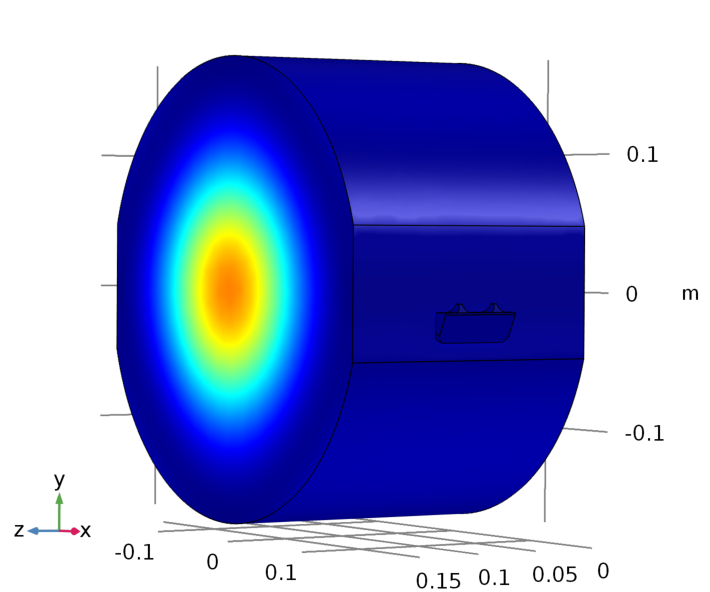} \vspace{-1mm}
\includegraphics[width=0.38\linewidth]{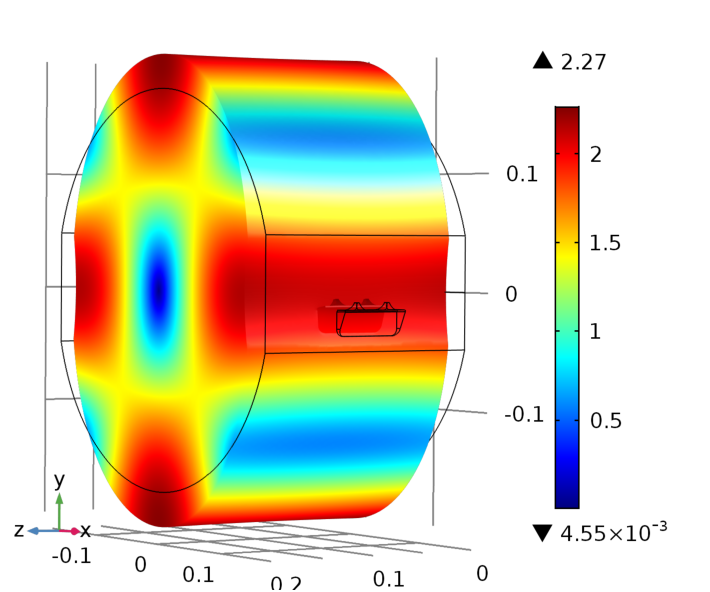} \vspace{-1mm}
\includegraphics[width=0.38\linewidth]{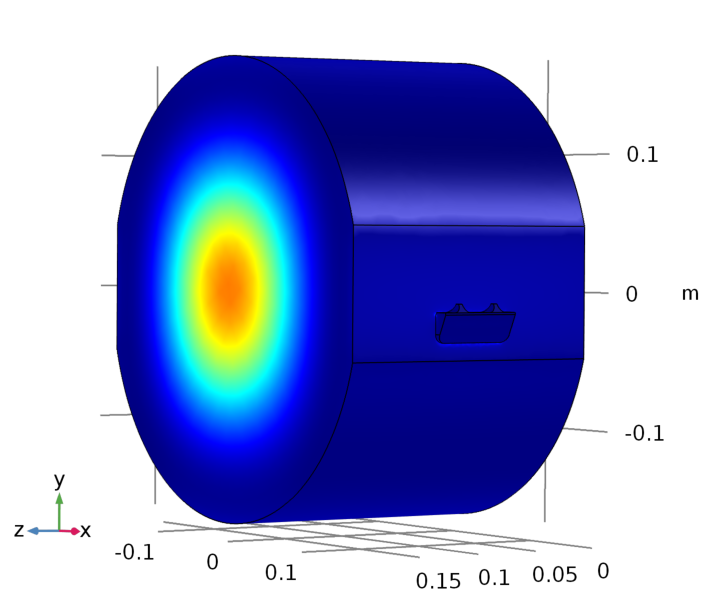} \vspace{-1mm}
\caption{Exaggerated mode displacements on the left and the form factors on the right.  These modes show how two eigenfrequencies can have very similar form factors.  They represent mode frequencies 8164 (top) and 8331\,Hz (bottom)} \label{fig:formfactors_similar}
\end{figure}

\begin{figure}[ht]
\centering
\includegraphics[width=0.43\linewidth]{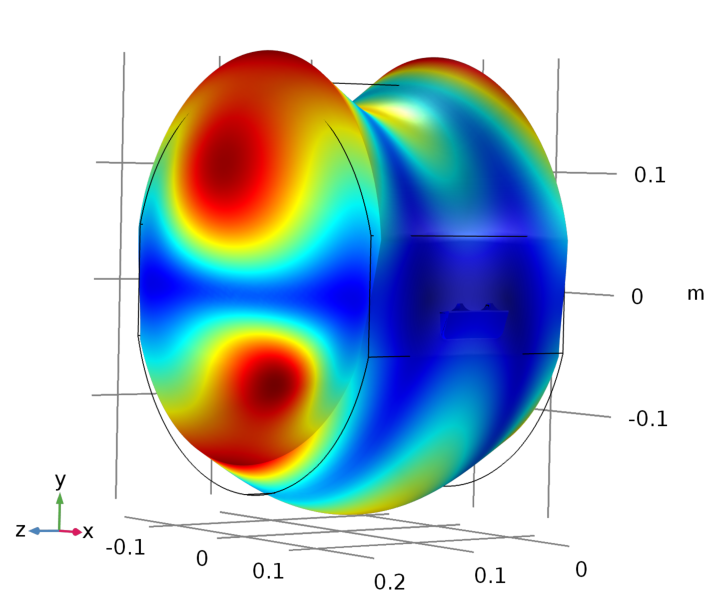} 
\includegraphics[width=0.43\linewidth]{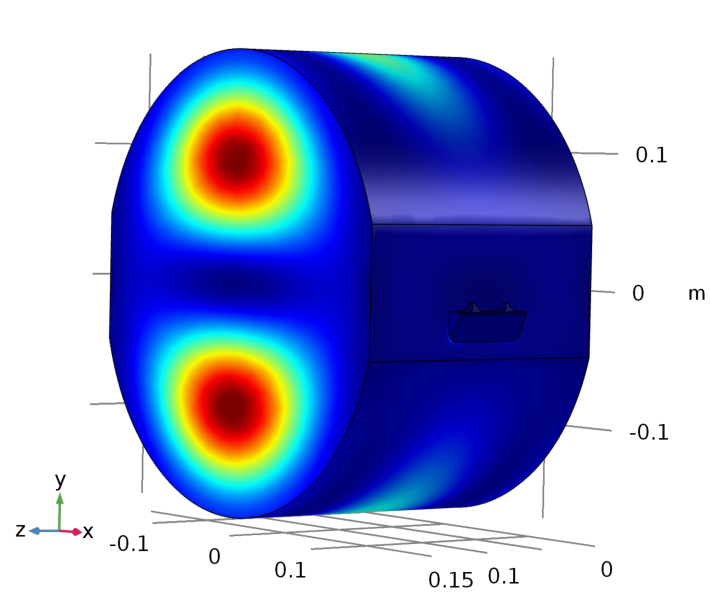} \vspace{-1mm}
\includegraphics[width=0.43\linewidth]{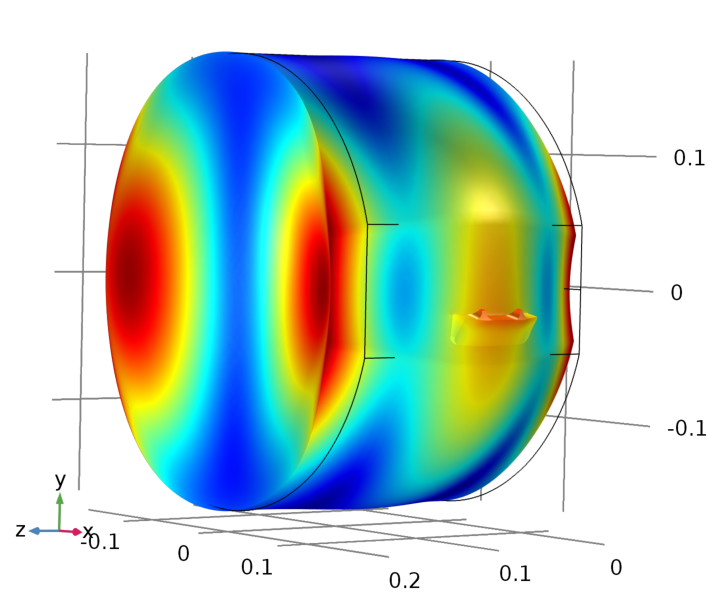} \vspace{-1mm}
\includegraphics[width=0.43\linewidth]{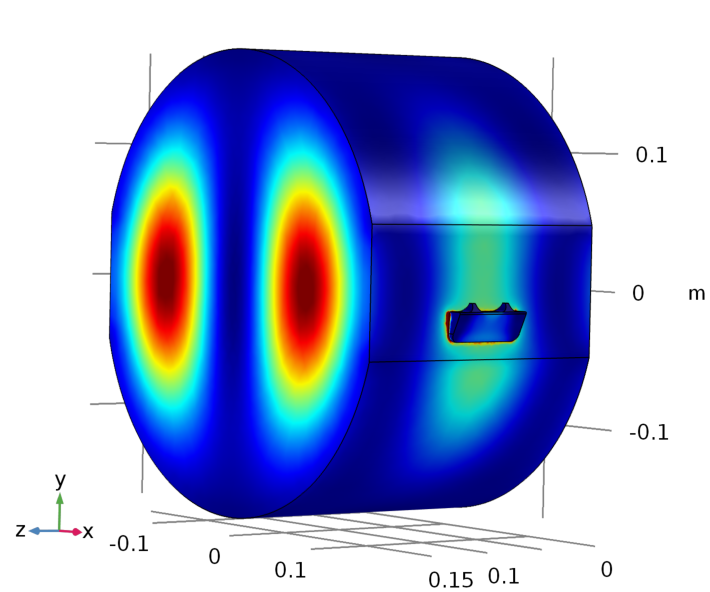} \vspace{-1mm}
\includegraphics[width=0.43\linewidth]{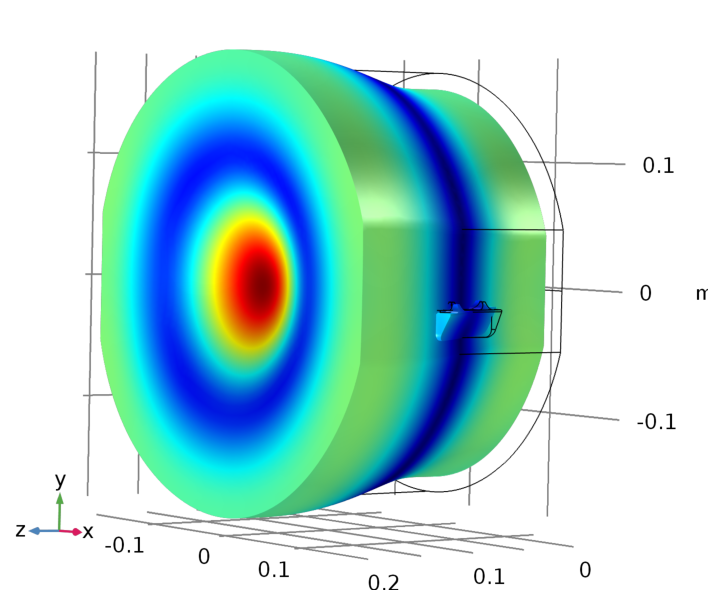} \vspace{-1mm}
\includegraphics[width=0.43\linewidth]{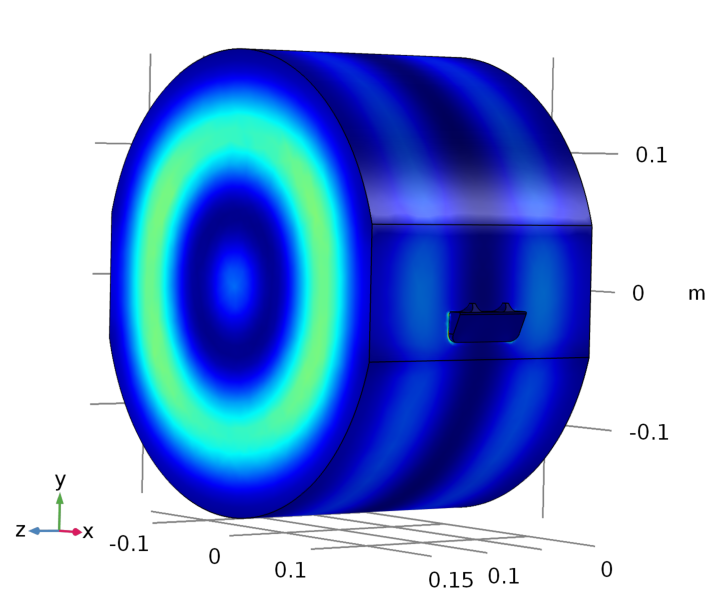} \vspace{-1mm}
\includegraphics[width=0.43\linewidth]{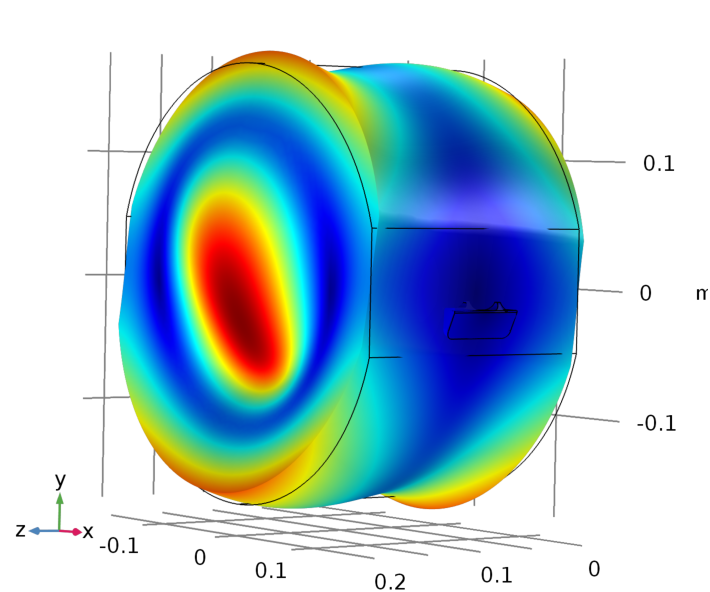} \vspace{-1mm}
\includegraphics[width=0.43\linewidth]{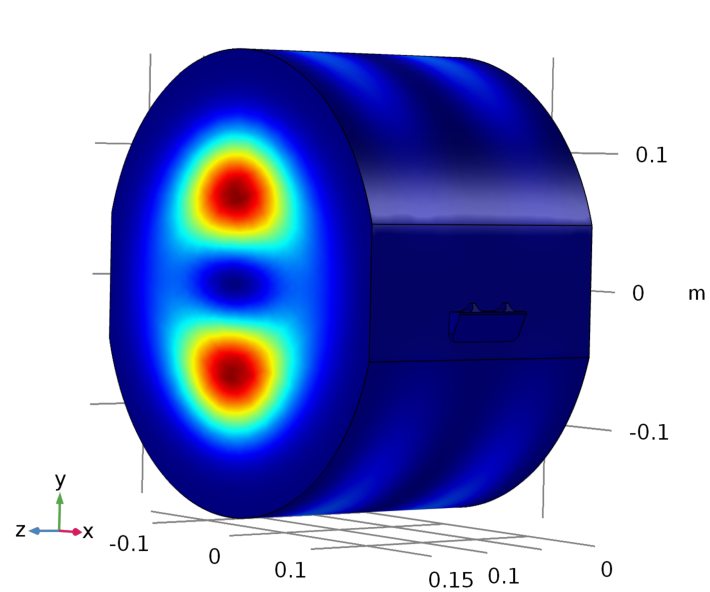} \vspace{-1mm}
\includegraphics[width=0.43\linewidth]{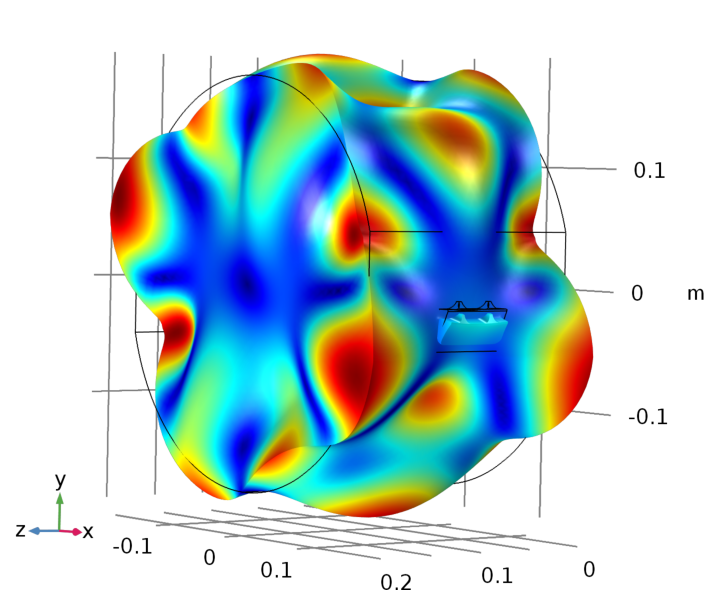} \vspace{-1mm}
\includegraphics[width=0.43\linewidth]{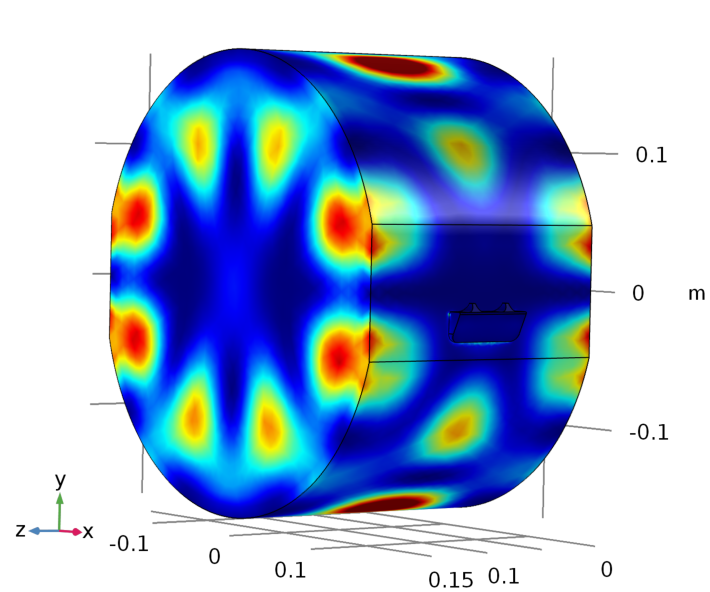} \vspace{-1mm}
\includegraphics[width=0.43\linewidth]{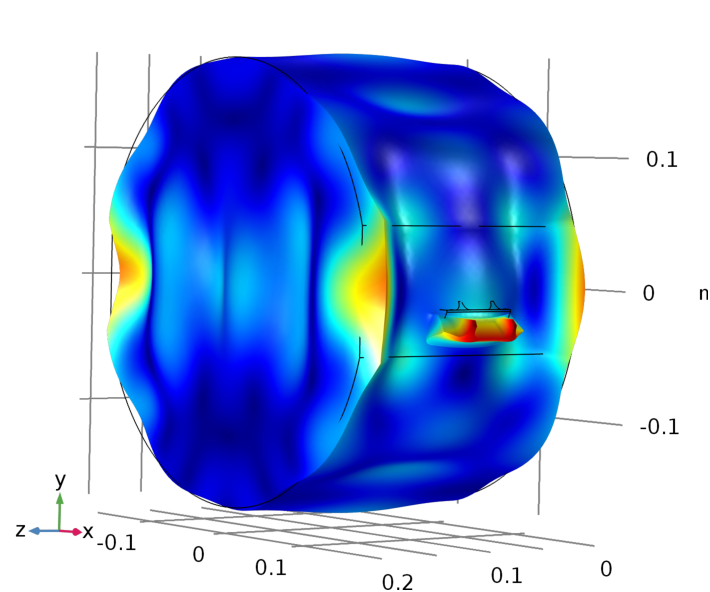} \vspace{-1mm}
\includegraphics[width=0.43\linewidth]{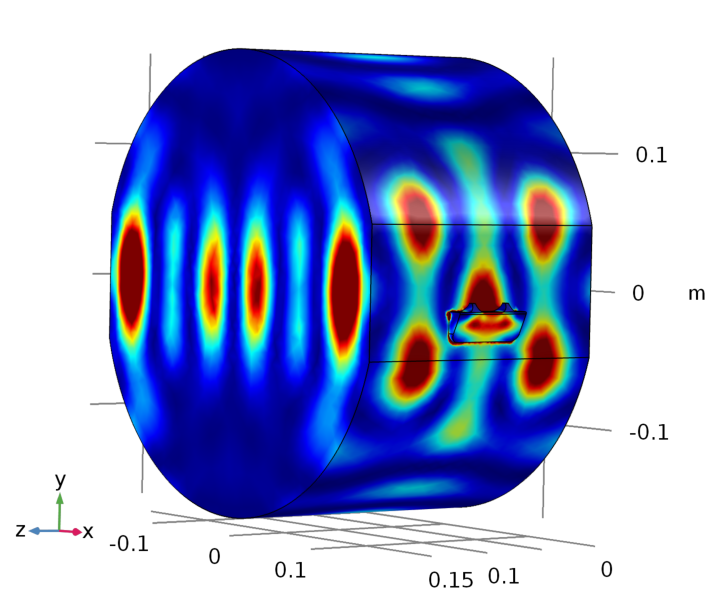} \vspace{-1mm}
\includegraphics[width=0.43\linewidth]{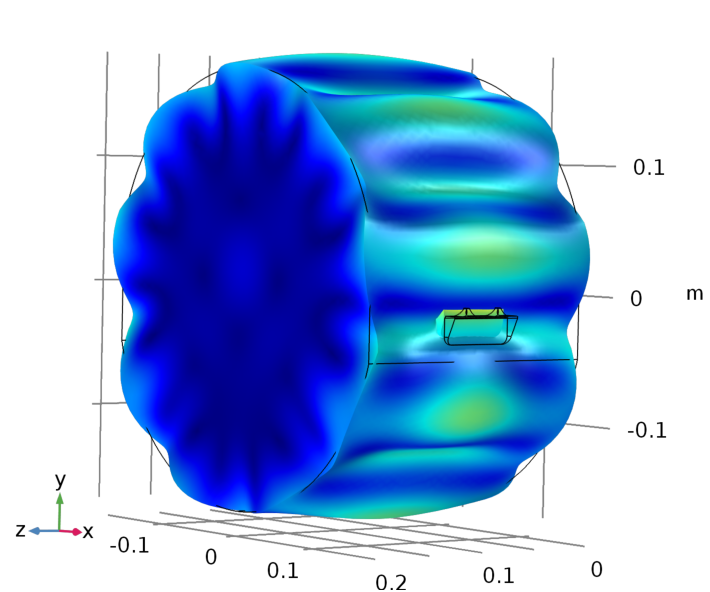} \vspace{-1mm}
\includegraphics[width=0.43\linewidth]{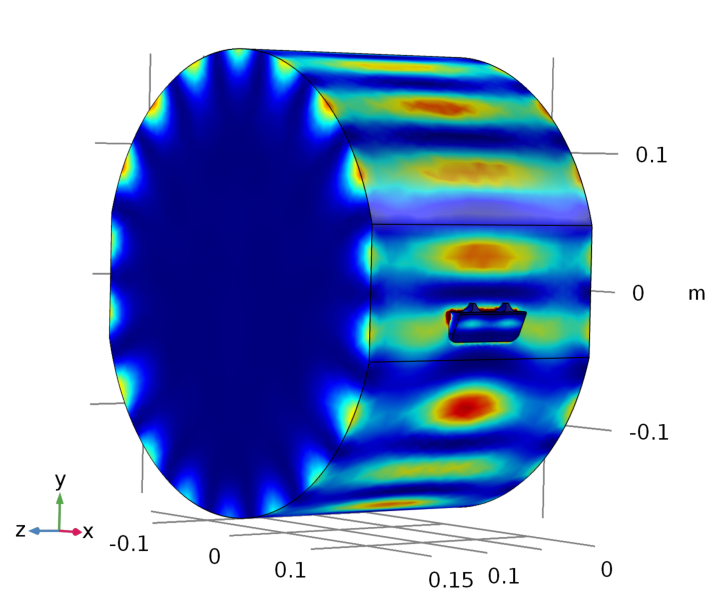} \vspace{-1mm}
\caption{The exaggerated mode displacements on the left and the form factors for these modes on the right.  These modes show a range of spatial scales attainable with existing LIGO measurements. They represent mode frequencies 15016, 15083, 15220, 15534, 23656, 33381 and 33610\,Hz (top to bottom)} \label{fig:formfactors_scale}
\end{figure}

\vspace{-2mm}
Figure~\ref{fig:formfactors_similar} shows a pair of eigenfrequencies that have very similar form factors.  These similarities make the conversion matrix rank deficient and thus singular value decomposition is required.   Figure~\ref{fig:formfactors_scale} shows a selection of vibrational wavefunctions and their associated form factors.  Blue regions of the form factors indicates areas of the test mass where the particular mode frequency is insensitive to temperature variation.  Red regions are areas where the mode frequency is sensitive to temperature variation.  These eigenfrequencies are measured in LIGO data and therefore formfactors represent temperature distribution spatial scale factors that should be measurable. 

\clearpage

\section{Singular Value Decomposition Eigenfunctions} \label{sec:SVDEle}

\begin{figure}[ht]
\centering
\includegraphics[width=0.9\linewidth]{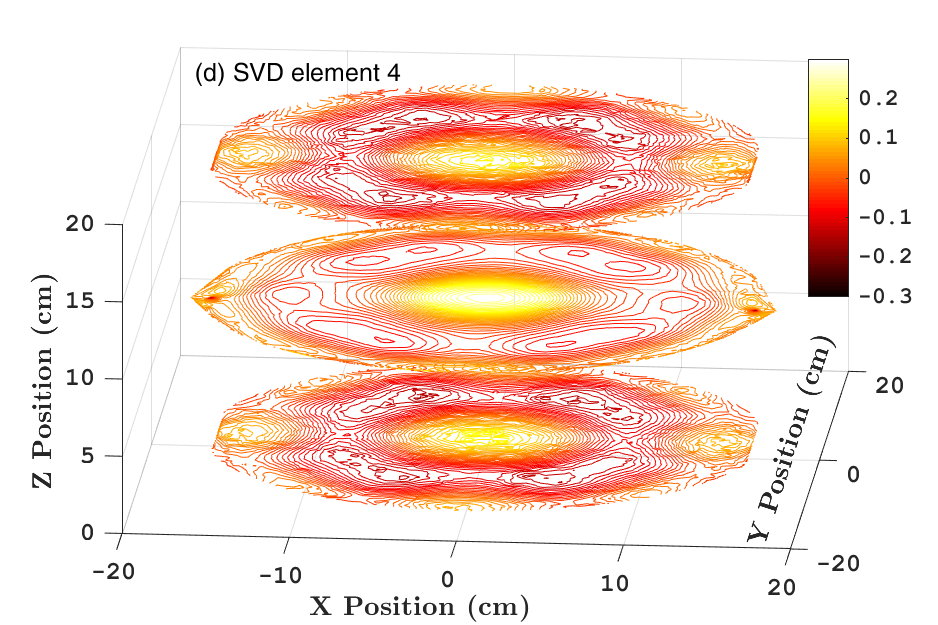} 
\includegraphics[width=0.9\linewidth]{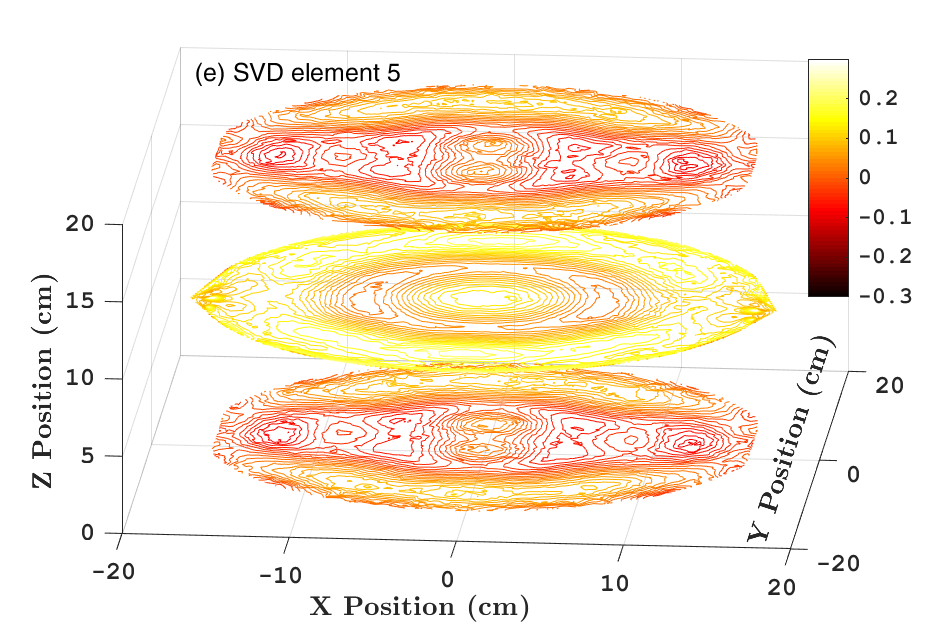} \vspace{-1mm}
\includegraphics[width=0.9\linewidth]{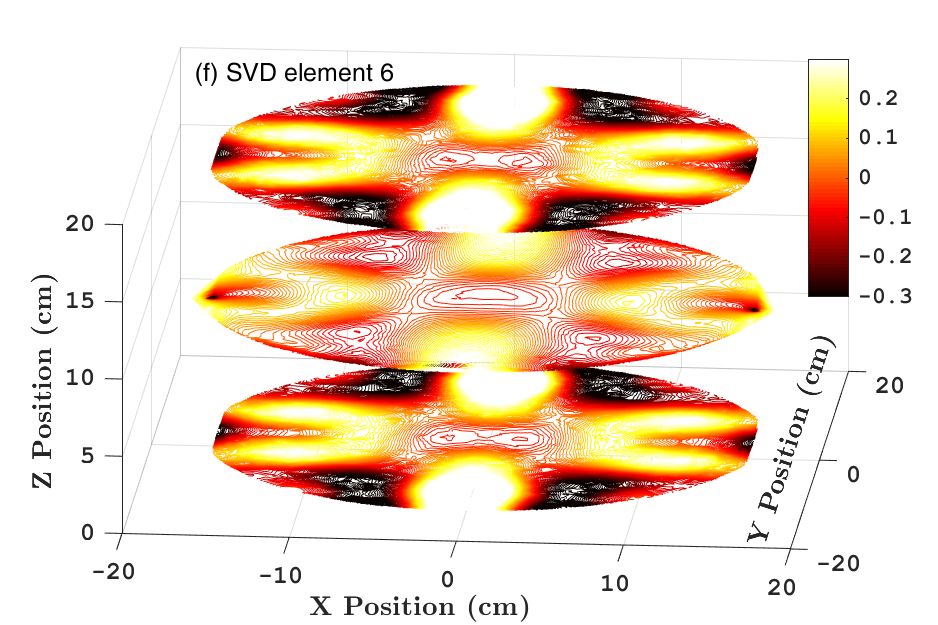} 
\includegraphics[width=0.9\linewidth]{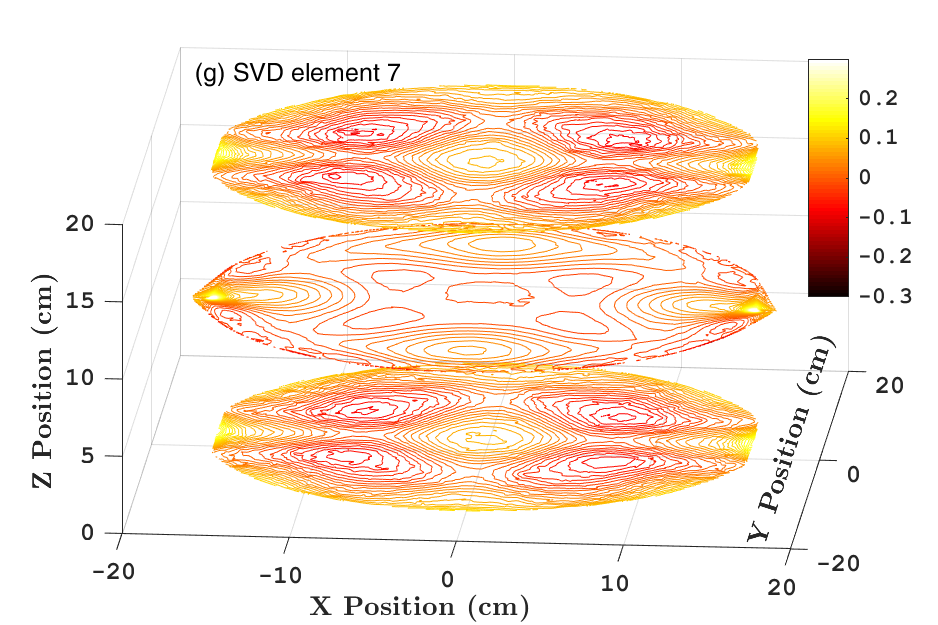} 
\caption{The seven largest eigenvalued eigenfunctions of the singular value decomposition described in section~\ref{sec:Tdist}. } \label{fig:svdElements}
\end{figure}

Figure~\ref{fig:svdElements} shows the first seven singular value decomposition element eigenfunctions for the example give in section~\ref{sec:Tdist}.  These are the elements with the largest eigenvalues and indicate the shapes of temperature distributions most easily recovered from eigenfrequency measurements.

\begin{figure}[ht!]
\centering
\includegraphics[width=0.9\linewidth]{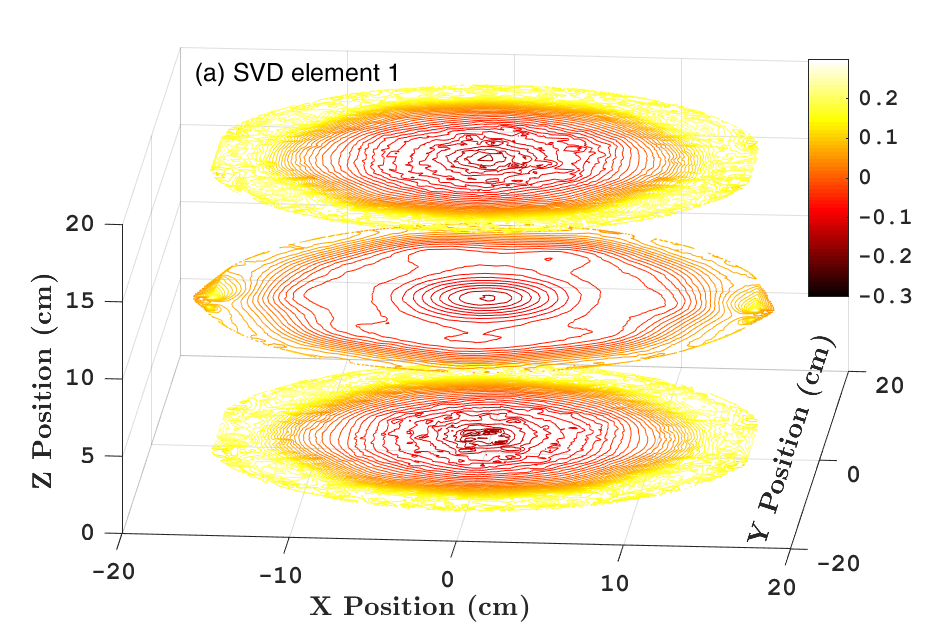} 
\includegraphics[width=0.9\linewidth]{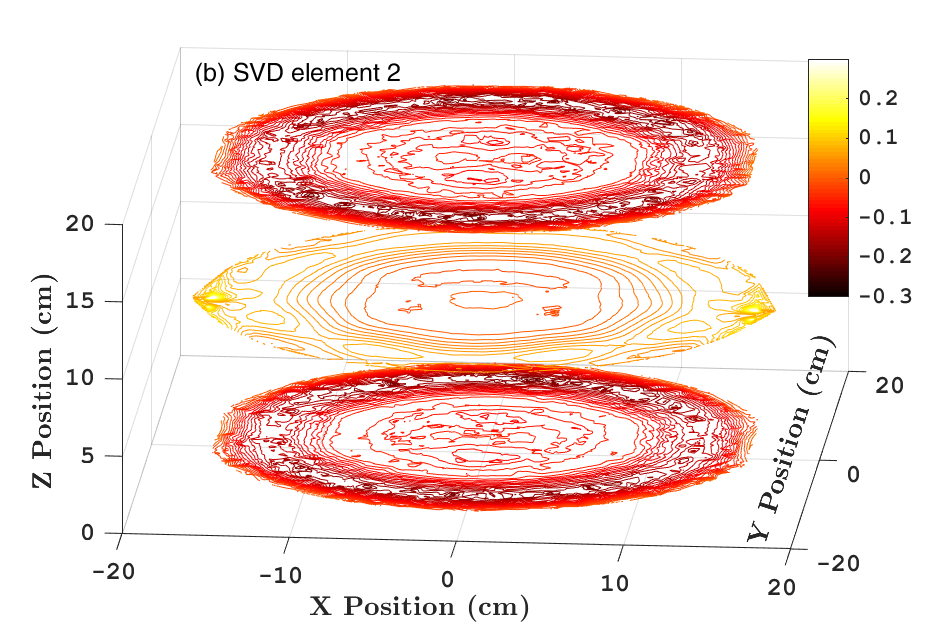} 
\includegraphics[width=0.9\linewidth]{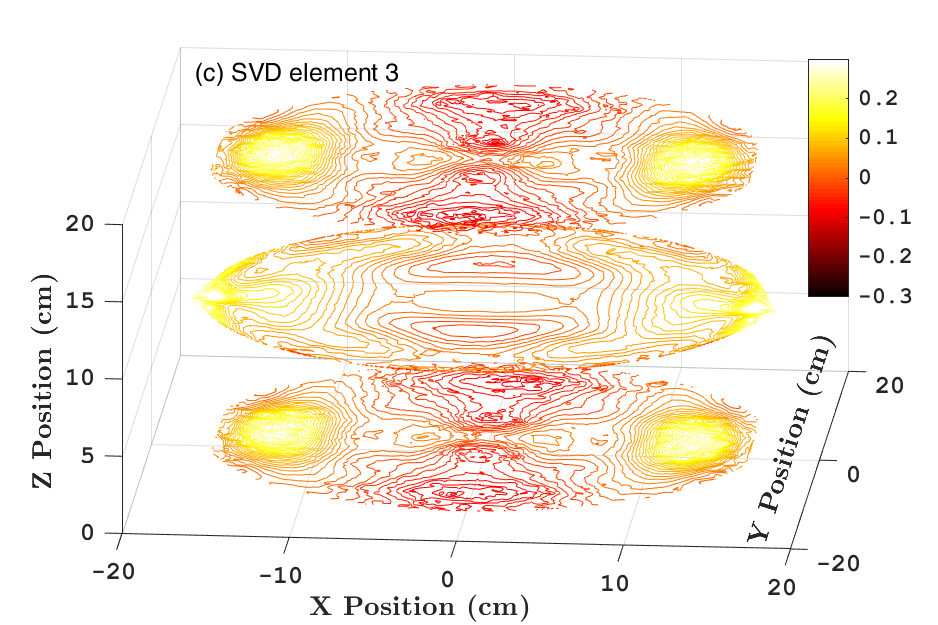} 
\end{figure}

\clearpage

\end{document}